\documentclass[12pt]{article}
\usepackage[margin=20mm,includefoot]{geometry}

\usepackage{cite}
\usepackage{amsmath, amssymb} 				
\usepackage{marvosym}

\numberwithin{equation}{section}

\usepackage{hyperref}
\hypersetup{
	unicode,	
	colorlinks,
	citecolor=cyan,
	linkcolor=[rgb]{0,.6,1},
	bookmarksnumbered,
	bookmarksopen=true,
	bookmarksopenlevel=\maxdimen,
	}

\def\vev#1{\langle #1\rangle}
\def\on#1#2{{\buildrel{\mkern2.5mu#1\mkern-2.5mu}\over{#2}}}
\def\dt#1{\on{\hbox{\bf .}}{#1}}                
\mathcode`\*="702A                  
\def\f#1#2{{\textstyle{#1\over#2}}}	   
\def\half{{\textstyle{1\over{\raise.1ex\hbox{$\scriptstyle{2}$}}}}}

\catcode128=13 \def €{\"A}                 
\catcode129=13 \def {\AA}                 
\catcode130=13 \def '{\c}           	   
\catcode131=13 \def ƒ{\'E}                   
\catcode132=13 \def "{\~N}                   
\catcode133=13 \def …{\"O}                 
\catcode134=13 \def †{\"U}                  
\catcode135=13 \def ‡{\'a}                  
\catcode136=13 \def ˆ{\`a}                   
\catcode137=13 \def ‰{\^a}                 
\catcode138=13 \def Š{\"a}                 
\catcode139=13 \def ‹{\~a}                   
\catcode140=13 \def Œ{\alpha}            
\catcode141=13 \def {\chi}                
\catcode142=13 \def Ž{\'e}                   
\catcode143=13 \def {\`e}                    
\catcode144=13 \def {\^e}                  
\catcode145=13 \def '{\"e}                
\catcode146=13 \def '{\'\i}                 
\catcode147=13 \def "{\`\i}                  
\catcode148=13 \def "{\^\i}                
\catcode149=13 \def •{\"\i}                
\catcode150=13 \def –{\~n}                  
\catcode151=13 \def —{\'o}                 
\catcode152=13 \def ˜{\`o}                  
\catcode153=13 \def ™{\^o}                
\catcode154=13 \def š{\"o}                 
\catcode155=13 \def ›{\~o}                  
\catcode156=13 \def œ{\'u}                  
\catcode157=13 \def {\`u}                  
\catcode158=13 \def ž{\^u}                
\catcode159=13 \def Ÿ{\"u}                
\catcode160=13 \def  {\tau}               
\catcode162=13 \def ¢{\oplus}           
\catcode163=13 \def £{\relax\ifmmode\expandafter\to\else\itemize\fi} 
\catcode164=13 \def ¤{\subset}	  
\catcode165=13 \def ¥{\infty}           
\catcode166=13 \def ¦{\mp}                
\catcode167=13 \def §{\sigma}           
\catcode168=13 \def ¨{\rho}               
\catcode169=13 \def ©{\gamma}         
\catcode170=13 \def ª{\leftrightarrow} 
\catcode171=13 \def «{\relax\ifmmode\expandafter\acute\else\expandafter\'\fi}
\catcode172=13 \def ¬{\relax\ifmmode\expandafter\ddt\else\expandafter\"\fi}
\catcode173=13 \def ­{\equiv}            
\catcode174=13 \def ®{{{}={}}}          
\catcode175=13 \def ¯{\Omega}          
\catcode176=13 \def °{\otimes}          
\catcode177=13 \def ±{\ne}                 
\catcode178=13 \def ²{\le}                   
\catcode179=13 \def ³{\ge}                  
\catcode180=13 \def ´{\upsilon}          
\catcode181=13 \def µ{\mu}                
\catcode182=13 \def ¶{\delta}             
\catcode183=13 \def ·{\epsilon}          
\catcode184=13 \def ¸{\Pi}                  
\catcode185=13 \def ¹{\pi}                  
\catcode186=13 \def º{\beta}               
\catcode187=13 \def »{\partial}           
\catcode188=13 \def ¼{\nobreak\ }       
\catcode189=13 \def ½{\zeta}               
\catcode190=13 \def ¾{\sim}                 
\catcode191=13 \def ¿{\omega}           
\catcode192=13 \def À{\dt}                     
\catcode193=13 \def Á{\gets}                
\catcode194=13 \def Â{\lambda}           
\catcode195=13 \def Ã{\nu}                   
\catcode196=13 \def Ä{\phi}                  
\catcode197=13 \def Å{\xi}                     
\catcode198=13 \def Æ{\psi}                  
\catcode199=13 \def Ç{\int}                    
\catcode200=13 \def È{\oint}                 
\catcode201=13 \def É{\relax\ifmmode\expandafter\cdot\else\vol\fi}    
\catcode202=13 \def Ê{\relax\ifmmode\expandafter\,\else\thinspace\fi}
\catcode203=13 \def Ë{\`A}                      
\catcode204=13 \def Ì{\~A}                      
\catcode205=13 \def Í{\~O}                      
\catcode206=13 \def Î{\Theta}              
\catcode207=13 \def Ï{\theta}               
\catcode208=13 \def Ð{\relax\ifmmode\expandafter\bar\else\expandafter\=\fi}
\catcode209=13 \def Ñ{\overline}             
\catcode210=13 \def Ò{\langle}               
\catcode211=13 \def Ó{\relax\ifmmode\expandafter\{\else\ital\fi}      
\catcode212=13 \def Ô{\rangle}               
\catcode213=13 \def Õ{\}}                        
\catcode214=13 \def Ö{\sla}                      
\catcode215=13 \def ×{\relax\ifmmode\expandafter\check\else\expandafter\v\fi}
\catcode216=13 \def Ø{\"y}                     
\catcode217=13 \def Ù{\"Y}  		    
\catcode218=13 \def Ú{\Leftarrow}       
\catcode219=13 \def Û{\Leftrightarrow}       
\catcode220=13 \def Ü{\relax\ifmmode\expandafter\Rightarrow\else\sect\fi}
\catcode221=13 \def Ý{\sum}                  
\catcode222=13 \def Þ{\prod}                 
\catcode223=13 \def ß{\widehat}              
\catcode224=13 \def à{\pm}                     
\catcode225=13 \def á{\nabla}                
\catcode226=13 \def â{\quad}                 
\catcode227=13 \def ã{\in}               	
\catcode228=13 \def ä{\star}      	      
\catcode229=13 \def å{\sqrt}                   
\catcode230=13 \def æ{\^E}			
\catcode231=13 \def ç{\Upsilon}              
\catcode232=13 \def è{\"E}    	   	 
\catcode233=13 \def é{\`E}               	  
\catcode234=13 \def ê{\Sigma}                
\catcode235=13 \def ë{\Delta}                 
\catcode236=13 \def ì{\Phi}                     
\catcode237=13 \def í{\`I}        		   
\catcode238=13 \def î{\iota}        	     
\catcode239=13 \def ï{\Psi}                     
\catcode240=13 \def ð{\times}                  
\catcode241=13 \def ñ{\Lambda}             
\catcode242=13 \def ò{\cdots}                
\catcode243=13 \def ó{\^U}			
\catcode244=13 \def ô{\`U}    	              
\catcode245=13 \def õ{\bo}                       
\catcode246=13 \def ö{\relax\ifmmode\expandafter\hat\else\expandafter\^\fi}
\catcode247=13 \def÷{\relax\ifmmode\expandafter\tilde\else\expandafter\~\fi}
\catcode248=13 \def ø{\ll}                         
\catcode249=13 \def ù{\gg}                       
\catcode250=13 \def ú{\eta}                      
\catcode251=13 \def û{\kappa}                  
\catcode252=13 \def ü{\half}     		 
\catcode253=13 \def ý{\Gamma} 		
\catcode254=13 \def þ{\Xi}   			
\catcode255=13 \def ÿ{\relax\ifmmode\expandafter{}^{\dagger}{}\else\dag\fi}

\def\rgboo#1{\pdfliteral{#1 rg #1 RG}}
\def\rgb#1#2{\rgboo{#1}#2\rgboo{0 0 0}}

\def\W{{}^\ast W}
\def\t#1{{#1}^\prime}

\title{
\Huge\bfseries\sffamily \strut\rgb{.9 0 1}{F-theory Superspace}
}
\author{
	William D.~Linch {\sc iii}$^{ \mbox{\footnotesize\Pisces}}$ 
	and
	Warren Siegel$^{ \mbox{\footnotesize\Scorpio}}$ 
	}
\date{}

\begin{document}
\hfill UMDEPP-015-001

\hfill YITP-SB-15-1

\vskip 1.2cm
\begin{center}
{\Huge\bfseries\sffamily \strut\rgb{.9 0 1}{F-theory Superspace} }\\
\vskip 0.7cm
{\large	William D.~Linch {\sc iii}$^{ \mbox{\footnotesize\Pisces}}$ 
	and
	Warren Siegel$^{ \mbox{\footnotesize\Scorpio}}$ }\\
\vskip 0.5cm
{\em
$^{ \mbox{\footnotesize\Pisces}}$Center for String and Particle Theory\\
Department of Physics,
University of Maryland at College Park,\\
College Park, MD 20742-4111 USA.\\
~\\
$^{ \mbox{\footnotesize\Scorpio}}$C. N. Yang Institute for Theoretical Physics\\
State University of New York, Stony Brook, NY 11794-3840
}
\end{center}

\vspace{10pt}

\begin{abstract}
We consider, at the linearized level, the superspace formulation of lower-dimensional F-theory.  In particular, we describe the embedding of 3D Type II supergravity of the superstring, or 4D, $N=1$ supergravity of M-theory, into the corresponding F-theory in full detail, giving the linearized action and gauge transformations in terms of the prepotential.  This manifestly supersymmetric formulation reveals some features not evident from a component treatment, such as Weyl and local S-supersymmetry invariances.  The linearized multiplet appears as a super 3-form (just as that for the manifestly T-dual theory is a super 2-form), reflecting the embedding of M-theory (as the T-dual theory embeds Type II supergravity).  We also give the embedding of matter multiplets into this superspace, and derive the F-constraint from the gauge invariance of the gauge invariance.
\end{abstract}

\vspace{1cm}
\begin{flushleft}
~\\
$^{\mbox{\footnotesize\Pisces}}$\href{mailto:wdlinch3@gmail.com}{wdlinch3@gmail.com}\\
$^{\mbox{\footnotesize\Scorpio}}$\href{mailto:siegel@insti.physics.sunysb.edu}{siegel@insti.physics.sunysb.edu}\\
\end{flushleft}

\setcounter{page}0
\thispagestyle{empty}

\newpage

\thispagestyle{empty}

{
\small
\tableofcontents}
\thispagestyle{empty}

\newpage

\section{Introduction}

\subsection{Earlier work}

After dimensional reduction only linear coordinate transformations are preserved, since those transform indices but not vanishing coordinates.  (There are also translations, but they don't affect indices.)  The vielbein in pure gravity (actually the compactification scalars) thus becomes an element of GL(D)/O(D$-$1,1), since the local tangent space symmetry is unaffected.  (If only d dimensions are compactiifed, replace D with d.)  This symmetry can be used to reconstruct gravity upon decompactifying the dimensions \cite{Ogievetsky:1978mt}.

If a similar reduction is applied to the vielbein and 2-form of the massless sector of the bosonic string, the fields are found to form the coset O(D,D)/O(D$-$1,1)$^2$ \cite{Duff:1989tf}, where O(D,D) is T-duality.  This symmetry can be used to reconstruct the complete low-energy limit of string theory in a manifestly T-duality covariant form by doubling the number of spacetime dimensions \cite{Siegel:1993xq, Siegel:1993th, Siegel:1993bj}:  The method is based on the string current algebra, which defines a new Lie derivative. The second D coordinates correspond to winding modes of the string; they do not require doubling of the string coordinates $X( ,§)$.  The coordinate transformations generated by the currents include both the usual coordinate transformations and the gauge transformations of the 2-form combined covariantly, and automatically imply the gauge invariance of the gauge invariance.  This allows for the definition of connections, covariant derivatives, torsions, curvatures, and Bianchi identities.  The action can then be constructed by introducing the dilaton as an O(D,D)-invariant integration measure (density).  An O(D,D)-covariant constraint (``strong constraint", as it was later called) is imposed on the coordinates for gauge invariance of the action; choice of a solution (``section") spontaneously breaks O(D,D), which is restored to O(d,d) upon reduction of d of the coordinates.  The construction generalizes straightforwardly to the superstring in superspace.  Recent developments include finite gauge transformations \cite{Hohm:2012gk} and the supersymmetric component action \cite{Hohm:2011nu,Jeon:2011sq}.  (Other recent developments are listed in \cite{Hohm:2013jaa}.)

Analogous cosets were found for dimensional reductions of M-theory \cite{West:2001as,Hull:2007zu}.  
Bosonic actions (without added coordinates) were constructed, but had measure problems, suggesting the procedure might work only for the complete 11D theory \cite{Berman:2010is}. Extra coordinates were found to manifest U-duality (including T-duality and S-duality), leading to a generalization of the strong constraint \cite{Berman:2011cg,Berman:2011jh,Berman:2011pe}.
We'll refer to these as ``{\bf F}-theory", according to its original definition as adding further dimensions to M-theory to incorporate U-duality.  (We'll also call this generalized constraint the ``F-constraint" to distinguish it from the original O(D,D) ``T-constraint".)  Supersymmetric generalizations were found, in components \cite{Godazgar:2014nqa, Musaev:2014lna}.  These field theory descriptions were related directly to branes in \cite{Hatsuda:2013dya,Hatsuda:2012vm,Hatsuda:2012uk}.

\subsection{Review}
\label{S:Review}
The most complete results were obtained for the massless bosonic sector of the {\bf F}-theory uplift of the 3D superstring \cite{Berman:2010is, Berman:2011cg, Park:2013gaj, Blair:2013gqa, Blair:2014zba}.  (The corresponding {\bf M}-theory is 4D, $N=1$ supergravity.)  The fully nonlinear gauge transformations and action were obtained (with the F-constraint).  The measure was not covariant, but the indications were that this was a consequence of not working in the critical dimension.  (In {\bf T}-theory the dilaton acts as the measure for all D.)  We will not see this problem here, as we limit ourselves to the linear approximation.  We now briefly summarize those results (with some reinterpretation).  

The symmetry unbroken by the vacuum is SO(3,1) for {\bf M} and SO(2,2) = SO(2,1)$^2$ for {\bf T} (left and right, doubling the SO(2,1) of the {\bf S}tring), so that for {\bf F} must be SO(3,2) = Sp(4).  This gives the usual naive extra dimension expected, but supersymmetry requires more:  The anticommutator of 2 supersymmetries yields for momentum a symmetric bispinor of Sp(4), i.e., an antisymmetric tensor {\bf 10} of SO(3,2).

From the usual dimensional reduction arguments (see more below) we know the bosons in this case must be an element of SL(5)/O(3,2).  But the vielbein must carry antisymmetric pairs of indices (as do the coordinates and momenta):  It is the {\bf 10} representation of the coset, while the defining {\bf 5} representation is more convenient, since it is unconstrained except for unit determinant (which we'll see below actually arises as a gauge invariance in the superspace formulation).  Expressing one in terms of the other,
$$ e_{ab}{}^{mn} = e_{[a}{}^m e_{b]}{}^n $$
where the indices belong to the {\bf 5} of SO(3,2) or SL(5).  Linearizing $e_a{}^m$ as symmetric, traceless {\bf 14} $h_{ab}$, the gauge transformations are thus
$$ ¶h_{ab} = »_{(a}{}^c Â_{b)c} - tr $$
($Â$ also a {\bf 10}).  

As for {\bf T}-theory, there is a gauge invariance of the gauge invariance, in this case
$$ ¶Â^{ab} = ü»_{cd}öÂ^{abcd} $$
in terms of the new parameter $öÂ^{abcd}$, a totally antisymmetric {\bf 5} of SO(3,2).  This requires for consistency the F-constraint
$$ »_{[ab}»_{cd]} = 0â(»_{a[b}»_{cd]} = 0) $$
acting on $öÂ^{abcd}$, and in turn on all other quantities.  At the nonlinear level, it is imposed even when the two derivatives act on different fields, since it must hold on products of functions (as required, e.g., for integration by parts).

These results relate to the usual for the {\bf M} and {\bf T}-theories:  {\bf M} comes from solving the constraint
$$ p^i_{[ab}p^j_{cd]} = 0 $$
for all momenta $p^i$, which implies
$$ p^i_{ab} = u_{[a}p^i_{b]}¼,â¶p^i_a ¾ u_a $$
Choosing a frame (and normalization) $u_a=¶_{-1,a}$ (for some time index ``$-1$"), we have only $p^i_{-1,a}$ nonvanishing, $a±-1$, a {\bf 4} of SO(3,1).  The bosonic sector of {\bf M} then follows straightforwardly (cf. appendix \ref{S:B}). The superspace reduction is worked out in section \ref{S:F2M}.

{\bf T} comes instead from dimensional reduction:  restricting the range of the indices $5£4$ (now dropping a space index as usual).  In this case $p$ becomes a {\bf 3$¢$3$'$} of SO(2,1)$^2$, the {\bf 3}'s of the 2 SO(2,1)'s coming from the selfdual and anti-selfdual parts of $p_{ab}$.  Also, $öÂ$ and the constraint reduce to the usual scalars, with the SO(2,2) inner product given by $·_{abcd}$.  The analysis is actually simpler in spinor notation (as implied by superspace) and is carried out in detail in section \ref{S:F2T}.

\subsection{Superspace}

In this paper we consider {\bf F}-theory as the U-duality completion of superstring theory in D = 3, 4, or 6, focusing mostly on D = 3 \cite{Polacek:2014cva}.  (The {\bf F}-theory itself is in much higher dimensions.)  We will study its formulation in superspace, but only at the linearized level.  Ultimately the goal is to use the richer yet more restrictive structure of superspace to help understand the geometry of the theory, and possibly the underlying worldvolume dynamics that implies it.  We'll see a few examples below.

We now consider general properties of supergravity in superspace for the related {\bf S}tring theory (Type II), manifestly {\bf T}-dual string theory, {\bf M}-theory, and {\bf F}-theory for cases corresponding to the (classical super)string theory in spacetime dimensions D = 3, 4, and 6.  {\bf M}-theory is defined in 1 dimension higher, {\bf T}-theory in double the dimensions, and {\bf F}-theory in whatever number needed to preserve the enlarged tangent-space symmetry, as described below.  The result can be summarized in table \ref{T:FMTS}.

\begin{table}[ht]
{\hfuzz=1000pt
$$
\vcenter{\openup.5\jot\halign{\hfil#\hfil&\quad\quad\quad\hfil#\hfil\cr
	G/H\span\omit\cr
	{\bf F}-theory\span\omit\cr
	bispinor coordinate\span\omit\cr
	\phantom{$\swarrow$}$\swarrow$&$\searrow$\phantom{$\searrow$}\cr
	GL(D+1)/O(D,1)&O(D,D)/O(D$-$1,1)$^2$\cr
	{\bf M}-theory& {\bf T}-dual theory\cr
	D+1&2D\cr
	\phantom{$\searrow$}\quad$\searrow$&$\swarrow$\quad\phantom{$\swarrow$}\cr
	GL(D)/O(D$-$1,1)\span\omit\cr
	{\bf S}tring (II)\span\omit\cr
	D\span\omit\cr}}
\qquad\qquad
\vcenter{\openup1\jot\halign{ # & \hfil#\hfil & \ \hfil#\hfil & \ #\hfil \cr
	D & G & H & bispinor (dimension)\cr
	3 & SL(5;\,$\mathbf R$) & Sp(4;\,$\mathbf R$) & symmetric (10)\cr
	4 & O(5,5) & Sp(4;\,$\mathbf C$) & hermitian (16)\cr
	6 & E$_{7(+7)}$ & SU*(8) & antisymmetric (28)\cr}
\vskip.4in\hskip-.5in $\swarrow$ sectioning \qquad $\searrow$ dimensional reduction} $$}
\caption{FMTS in superspace}
\label{T:FMTS}
\end{table}

The tangent-space symmetry H listed above for {\bf F}-theory follows from taking the covering group of SO(D$-$1,1) and doubling its argument.  The supersymmetry algebra (or that of the corresponding flat-space covariant derivatives) thus takes the same form as for the standard superspace for Type I supersymmetry,
\begin{align}
D=3: &âÓ d_Œ , d_º Õ = 2p_{Œº}¼,âp_{ºŒ} = p_{Œº} \nonumber \\
D=4: &âÓ d_Œ , Ðd_{Àº} Õ = p_{ŒÀº}¼,âÓ d , d Õ = Ó Ðd , Ðd Õ = 0 \nonumber \\
D=6: &âÓ d_Œ , Ðd_º Õ = p_{Œº}¼,âÓ d , d Õ = Ó Ðd , Ðd Õ = 0¼,âp_{ºŒ} = - p_{Œº}
\end{align}
except for this doubling:  The fermionic coordinates are in the defining (``spinor") representation of H, and the bosonic coordinates are in the bispinor (``vector") representation with the same symmetry/reality conditions.  The number of supersymmetries is thus the same as for {\bf S}-theory (Type II, as well as for {\bf M} and {\bf T}).  However, the bosonic space is enlarged due to the larger H.

The cosets G/H give all of the bosonic sector of {\bf F}.  (The cosets given for {\bf M} and {\bf S} give just gravity, while that for {\bf T} gives only gravity and the 2-form.)  They correspond to the Lorentz scalars of maximal supergravity in 10$-$D dimensions (except that H is Wick rotated for 2 time dimensions \cite{Hull:1998br}), while the bispinor coordinates correspond to the Lorentz vectors (which are representations of G).  The ``vector" vielbein is in the coset G/H, in the bispinor representation, which can be expressed as a homogeneous polynomial in that of the defining representation.

The D = 3 case might better be considered as GL(5)/SO(3,2)$°$GL(1), indicating gravity in 3+2 dimensions with local Weyl scale invariance.  (Supergravity then has also local S-supersymmetry, as we'll see below.)  The case D = 4 resembles a T-dual theory for 5D supergravity, except for the reality conditions and the mixed spinor indices of the bosonic coordinates.  For D = 6 note that, unlike SU(8), all even rank tensors of SU*(8) including the bispinor are real, since the defining representation is pseudoreal.   Furthermore, SU*(8) has subalgebras SO*(8) = SO(6,2) and SU*(4)$^2$ = SO(5,1)$^2$.  But SU*(8) = SL(4;$\mathbf H$), not Sp(4;$\mathbf H$).
For the rest of the paper we concentrate on D¼=¼3.

\subsection{Super F-orms}
The flat space covariant derivative algebra is 
\begin{align}
\label{E:SUSYalgebra}
\{ d_\alpha, d_\beta \} = 2 p_{\alpha \beta}
\end{align}
where $p_{\alpha \beta}= p_{\beta \alpha}$ is {symmetric}. 
For D=3, we can write 
\begin{align}
p_{\alpha \beta} = \tfrac12 (\gamma^{ab})_{\alpha\beta} p_{a b}¼,âp_{ab}= -p_{ba},
\end{align} 
where $p$ is antisymmetric in vector indices. This superspace has 10 bosonic coordinates (with signature (6,4) cf. \S \ref{S:Reduction}) and $4$ fermionic ones.

For gauge invariance, we require the ``F-constraint''
\begin{align}
\label{E:Fconstraint}
p^2_{\vev{\alpha \beta}} = 0
~~~\Leftrightarrow~~~
\epsilon^{abcde}p_{bc}p_{de} =0 
\end{align}
on all fields and gauge parameters. Here, $X_{\vev{\alpha \beta}} := X_{[\alpha \beta]} + \frac12C_{\alpha \beta} X^\gamma{}_\gamma$ stands for the anti-symmetric part with the $C$-trace removed.
We will at times refer to {\bf F} as ten-dimensional but the theories are effectively restricted to satisfy this constraint. Then, from (\ref{E:d4b}, \ref{E:d5}) 
we obtain
\begin{align}
\label{dred}
d_{\langle \alpha} d^2 d_{\beta\rangle} = 0
~~~\mathrm{and}~~~ 
d^2 d_\alpha d^2 = 0.
\end{align}
Starting in the next section, we will derive the {\bf F}-analogues of the familiar matter, gauge, and gravitational multiplets. Together, they form a de Rham-type hierarchy which we represent in table \ref{T:F} (cf. \cite{Gates:1980ay,Gates:1983nr}). 

The F-constraint gives rise to a new type of bosonic representation superficially similar to exterior differential forms in ten dimensions but with the opposite parity on its indices. In contrast to the ten-dimensional de Rham differential $d =\tfrac12 dx^{ab} \partial_{ab}$, which squares to zero by commutativity of $\partial$, these forms are solutions of $\partial_{[ab}\omega_{c_1\dots c_p]}=0$ of the form $\omega_{c_1\dots c_p} = \partial_{[c_1c_2} \eta_{c_3\dots c_p]}$ which is a valid solution due to the vanishing of $\partial_{[ab} \partial_{cd]} = \partial_{[cd} \partial_{ab]}$ by the F-constraint (\ref{E:Fconstraint}). As we will see in section \ref{S:Matter}, these ``{\bf F}-orms'' appear in the component expansions of the {\bf F}-theory superfields and reduce to $p$-forms under sectioning {\bf F} $\to$ {\bf M}. 
They are worked out in appendix \ref{S:B} where it is found that they come in two distinct de Rham-type sequences we will call $p$-forms and $\ast p$-forms based on the degree of the {\bf M}-theory $p$-form they describe (cf.~tables \ref{E:p} and \ref{E:*p}).


\begin{table}[ht]
$$\vcenter{\tabskip=0pt \offinterlineskip
\halign{\strut#& \vrule#\tabskip=1em plus2em& \hfil$#$\hfil& \vrule#& \hfil$#$\hfil&
	 \vrule#& \hfil$#$\hfil& \vrule#& \hfil#\hfil& \vrule#& \hfil$#$\hfil& \vrule#\tabskip=0pt\cr
	\noalign{\hrule}
        \vrule height13pt depth3.5pt width0pt && p¼\hbox{(mod 6)} && ßA_p && \widehat{dA}_p && multiplet && \hbox{EOM} &\cr
	\noalign{\hrule}
        \vrule height13pt depth3.5pt width0pt && 0 && A && d^2 A && scalar (\S\ref{S:SM}) && d^2\widehat{dA} &\cr
        \vrule height13pt depth3.5pt width0pt && 1 && A && d^2 d_Œ A && vector (\S\ref{S:VM}) && d^Œ\widehat{dA}_Œ &\cr
        \vrule height13pt depth3.5pt width0pt && 2 && A_Œ && d_{Ҍ}A_{ºÔ} && superconformal (\S\ref{S:Superconformal}) && d^º\widehat{dA}_{Œº} &\cr
        \vrule height13pt depth3.5pt width0pt && 3 && A_{ҌºÔ} && d^4 A_{ҌºÔ}+... && supergravity (\S\ref{S:SGAction}) && \widehat{dA}_{Œº} &\cr\vrule height13pt depth3.5pt width0pt && 4 && A_{ҌºÔ} && d^º A_{ҌºÔ} && dual vector (\S\ref{S:dVM}) && d_{Ҍ}d^2\widehat{dA}_{ºÔ} &\cr
        \vrule height13pt depth3.5pt width0pt && 5 && A_Œ && d^Œ d^2 A_Œ && tensor (\S\ref{S:TM}) && d^2 d^Œ\widehat{dA} &\cr
	\noalign{\hrule}}}$$
\begin{align*}
\hbox{gauge transformation:} & â¶ßA_p = \widehat{dñ}_{p-1}  \\
\hbox{field strength:} & âßF_{p+1} = \widehat{dA}_p  \\
\hbox{Bianchi identity:} & â0 = \widehat{dF}_{p+1}
\end{align*}
\caption{Super {\bf F}-orms}
\label{T:F}
\end{table}

\subsection{Outline}

In section \ref{S:Matter}, we will construct each of the matter and gauge multiplets (corresponding to $p=0,1,4,5$ in table \ref{T:F}). 
The constraints defining the multiplets are complicated due to the lack of chirality. 
In particular, products of irreducible superfields are not irreducible superfields ({\it i.e.}~there is no analogue of the chiral ring).
It will turn out that every multiplet contains a bosonic $p$-form with a gauge transformation.
For each multiplet, we give the component structure and quadratic action. 
We then describe the (lack of) self-couplings and attempt to classify interactions between the multiplets. 
Such terms are severely restricted by gauge invariance and their complicated constraint structure. 

The result of the analysis of interactions is that all interactions between the various irreducible representations are at most quadratic with the scalar multiplet (SM) coupling to itself to give a mass term. The tensor multiplet (TM) and vector multiplets (VM) can only couple to each other in a $BF$-type interaction, thereby giving a mass to the TM. The dual vector multiplet (dVM) cannot couple to any of the other superfields. 

The field strength of the VM is the gauge transformation of the next entry in the de Rham complex. Instead of describing a 2-form (which already sits in the TM), it turns out that the associated invariant is the multiplet of linearized local superconformal transformations, which we study in section \ref{S:Fgravity}. The next entry of the table has this multiplet as a set of gauge transformations and, therefore, describes the ``conformal" graviton:
In the same way that the orthogonality of the vielbein in {\bf T}-theory makes its linearization a super 2-form, the linearized superfields of {\bf F}-supergravity are naturally associated with a super 3-form.
We construct its linearized Einstein tensor and, thus, the linearized supergravity action, invariant under the linearized superconformal transformations. We find, however, that we are unable to construct the super-Weyl tensor (as expected from a similar analysis at the nonlinear level in {\bf T}-theory).  

The derivation of {\bf M}, {\bf T}, and {\bf S}-theories by combinations of sectioning and dimensional reduction of all the above {\bf F}-theories is discussed in section \ref{S:Reduction}.  The final section contains our conclusions.

Some superspace $d$-identities are derived in appendix \ref{S:A}. In appendix \ref{S:B} we present a bosonic version of {\bf F}-theory. First, we address the subtle notion of $p$-forms which appear in the components of the matter and gauge multiplets in the main body of the paper. We end the appendices with the construction of bosonic {\bf F}-gravity.

\section{Matter}
\label{S:Matter}

\subsection{Scalar Multiplet}
\label{S:SM}
Consider the real scalar superfield constrained by
\begin{align}
\label{scalarBI}
d^2 d_\alpha {\phi} = 0.
\end{align}
Due to the identity (\ref{dred}), we can solve this constraint in terms of a (real) prepotential ${U}$ as\footnote{Note that this implies the existence of an $¥$ tower of multiplets, starting with the dual (tensor) multiplet (cf. \S \ref{S:TM})}
\begin{align}
\label{vprepot}
{\phi} =  d^2 {U},
~~~
\delta {U} = d^\alpha d^2 \xi_\alpha .
\end{align}
The $8+8$ components of the real scalar superfield ${U}$ are given by
\begin{align}
\label{v}
{U} \sim   {U} + \theta\eta + \theta^2\phi + \theta \gamma^a\theta A_a + \theta^2 \theta \psi + \theta^4 D.
\end{align}

We take the action to be
\begin{align}
\label{ScalarAction}
S = \frac 1{2} \int d^{4} x \int d^4 \theta \,  {\phi}^2 .
\end{align}
This yields the equation of motion
\begin{align}
d^2 {\phi} {{}={}} 0 .
\end{align}
Together with the constraint (\ref{scalarBI}), this implies the Dirac equation $p_{\alpha \beta}d^\beta {\phi} {{}={}} 0$ by the second identity in (\ref{dalgebra}).

The components of the scalar multiplet are
\begin{align}
{\phi} \sim {{\phi}} + \theta \psi  + \theta \gamma^a \theta F_a + \theta^2 D
~~~\mathrm{with}~~~
F_a = \partial_{ab} A^b
\end{align}
The ``vector'' component $F_a$ is a derivative of the vector $A^a$ in the prepotential ${U}$ (\ref{vprepot}) by the fifth and sixth identities in (\ref{dalgebra}). By the defining constraint (\ref{scalarBI}) and the first equation in (\ref{E:d4b}), it satisfies the Bianchi identity $\partial_{[ab} F_{c]} =0$ so that $F$ is a $\ast 2$-form (cf.~table \ref{E:*p}). Taken together, this multiplet carries $4+4$ off-shell degrees of freedom.
On-shell, the vector potential is further constrained to satisfy $0 = \Delta_a d^2 {\phi} \propto \partial_{ab} \Delta^b {\phi} \propto \partial_{ab}\partial^{bc}A_c$. This is the Maxwell-type wave equation $\partial^{ab}F_b=0$ (cf.~eq.~\ref{E:*p}) reducing the multiplet to $2+2$ on-shell degrees of freedom.

Reducing {\bf F} $\to$ {\bf M}, $d^2 \to d^2 + \bar d^2$ and the scalar superfield ${\phi} \to  \Pi + \bar \Pi$ reduces to a chiral field $\Pi$ with real prepotential plus its conjugate. 
This is the known description of the four-dimensional gauge 3-form multiplet \cite{Gates:1983nr}. We conclude that this ``scalar'' multiplet describes the {\bf F}-theory lift of the gauge 3-form.
We will revisit this reduction in section \ref{S:FtoM}.

\subsubsection{Scalar Multiplet Mass} It is impossible to write a general superpotential for this multiplet since, for example, in the reduction {\bf F} $\to$ {\bf M}, the resulting superpotential fails to be holomorphic.
The exception to this is the mass term
\begin{align}
S_m = \frac m{2} \int d^{4} x \int d^4 \theta \, {U} {\phi} 
,
\end{align}
which is gauge invariant by the constraint on ${\phi}$.
It reduces correctly to {\bf M} by the linearity in ${\phi}$:
\begin{align}
S_m &\to \frac m{2} \int d^{4} x \int d^4 \theta \, {U} (\Pi + \bar \Pi) 
	=\frac m{2} \int d^{4} x \int d^4 \theta \, {U} \Pi + \mathrm{h.c.}\cr
	&=\frac m{2} \int d^{4} x \int d^2 \theta \, \bar d^2 {U} \Pi + \mathrm{h.c.}
	=\frac m{2} \int d^{4} x \int d^2 \theta \, \Pi^2 + \mathrm{h.c.}
\end{align}
The massive SM equation of motion becomes
\begin{align}
 d^2 {\phi} + m {\phi}{{}={}} 0.
\end{align}

Note that this equation is annihilated by $d^2d_\alpha$. 
This requirement gives a stronger reason that there can be no superpotential $V({{\phi}})$  for the scalar multiplet: Consider the putative equation of motion $d^2 {{\phi}} - f({{\phi}}){{}={}} 0$ where $f= - V^\prime$. Acting with $d^2 d_\alpha$ automatically annihilates the first term. On the potential it gives
\begin{align}
0& {{}={}} 2d^2 d_\alpha f({{\phi}}) 
	= f^{\prime \prime \prime} d^\beta {{\phi}} d_\beta {{\phi}} d_\alpha {{\phi}}
		+ f^{\prime \prime} \left( 2d^2 {{\phi}} d_\alpha {{\phi}} + 2 d^\beta {{\phi}} d_\beta d_\alpha {{\phi}}\right)
\cr&
	{{}={}} \left( f^{\prime \prime \prime} d^\beta {{\phi}} d_\beta {{\phi}} 
		+ f^{\prime \prime}f \right) d_\alpha {{\phi}} + f^{\prime \prime} d^\beta {{\phi}} d_\beta d_\alpha {{\phi}} .
\end{align}
Canceling the second term requires $f^{\prime \prime} = 0$ which then implies that $f^{\prime\prime \prime} = 0$ as well. Thus, we recover that 
\begin{align}
f = \Lambda + m {{\phi}}
\end{align}
for some constant parameters $\Lambda$ and $m$ with mass dimensions $2$ and $1$, respectively.

\subsection{Tensor Multiplet}
\label{S:TM}
The dual ${G}$ of the scalar multiplet has its defining constraint and equation of motion reversed as compared to the equations defining the dynamical scalar multiplet ${\phi}$. 
That is, the constraint is 
\begin{align}
d^2 {G} = 0 
~~~\Rightarrow~~~
{G} =  d^\alpha d^2 \eta_\alpha
~:~
\delta \eta_\alpha = d^\beta \upsilon_{\alpha \beta}
\end{align}
for some unconstrained spinor superfield $\eta$ with gauge parameter $\upsilon_{\alpha \beta}=\tfrac12 \upsilon_{\vev{\alpha \beta}}$.

The action dual to the scalar action (\ref{ScalarAction}) is
\begin{align}
\label{E:TMAction}
S = -\frac 1{2} \int d^{4} x \int d^4 \theta \, {G}^2 .
\end{align}
The equation of motion extremizing this action is
\begin{align}
\label{eq:TMEoM}
d^2d_\alpha {G} {{}={}} 0 .
\end{align}
The Dirac equation on the spinor in ${G}$ is implied by the constraint and the equation of motion as with the scalar multiplet to which it is dual.

The components of the tensor multiplet are
\begin{align}
{G} \sim  {G} + \theta \tilde \psi + \theta \gamma^a \theta \tilde F_a .
\end{align}
The vector component is constrained by $\partial^{ab}\tilde F_{b} = \partial^{ab} \Delta_b {G} \propto \Delta^a d^2 {G} = 0$ as expected from duality with the scalar multiplet. This constraint identifies $\tilde F_a= \tfrac1{4!}\epsilon_a{}^{bcde}F_{bcde}$ as the (dual of the) field strength $F_{abcd} = \partial_{[ab} A_{cd]}$ of a 2-form (cf.~table \ref{E:p}). The equation of motion now implies $\partial^{cd} F_{abcd} =0$ reducing the $4+4$ component off-shell description to $2+2$ on-shell components.

\subsubsection{TM \texorpdfstring{$\leftrightarrow$}{\textleftarrow\textrightarrow} SM Duality} 
This multiplet can be obtained by performing a duality transformation on the scalar multiplet of the previous section. A duality functional for 2-form $\leftrightarrow \ast 2$-form is 
\begin{align}
\int d^{4} x \int d^4 \theta \, \left( {G} d^2 {U} - \tfrac12 {G}^2 \right) .
\end{align}
Indeed, integrating out ${G} $ gives ${G} = d^2 {U}$. Substituting back into the action gives the scalar multiplet of section \ref{S:SM}. Alternatively, integrating out ${U}$ gives the constraint $d^2 {G} =0 $ which is satisfied by the tensor multiplet solution ${G} = d^\alpha d^2 \eta_\alpha$, and the action reduced to the tensor muliplet action (\ref{E:TMAction}).

As usual, there is a ``dual" action for expressing this duality,
\begin{align}
Çd^4 xÊd^4 ϼ\left(Ä d^Œ d^2 \eta_Œ - üÄ^2\right) .
\end{align}
Now $Ä$ is auxiliary, and $G=d^Œ d^2 \eta_Œ$ appears explicitly.

\subsubsection{Tensor Multiplet Mass}
The equation of motion (\ref{eq:TMEoM}) is annihilated by contraction with $(d\gamma_a)^\alpha$. Since no combination of $m$, $d$, and $\eta$ has this property, we conclude that this multiplet cannot have a mass term. Indeed, postulating a superpotential $V_\alpha({G})$ and repeating the analysis of the previous section, we find that $d{G} \gamma_a f^\prime {{}={}} 0$, which is easily shown to imply that $f^\prime_\alpha =0$. Thus, $f_\alpha =0$ unless we consider couplings to other multiplets.
In section \ref{S:VM}, we will find that the vector multiplet field strength $W_\alpha$ satisfies this constraint. There we will take the constant (in ${G}$) $f_\alpha = {c_{BF}} W_\alpha$ which deforms the equation of motion to $d^2 d_\alpha {G} + m W_\alpha {{}={}} 0$ with $m={c_{BF}}$. 

\subsection{Vector Multiplet}
\label{S:VM}
The scalar and tensor multiplets of the previous section fit into a de Rham complex $\dots \to \mathrm{TM} \to \mathrm{SM} \to \dots$ in which the field ${G} = d^\alpha d^2\eta_\alpha$ in the TM becomes the gauge transformation of the prepotential ${U}$ of the SM $\phi =d^2{U}$. In this section we extend this complex to the right 
\begin{align}
\label{E:dR}
\dots \longrightarrow \mathrm{TM} \longrightarrow \mathrm{SM} \longrightarrow \mathrm{VM}\longrightarrow \dots
\end{align}
by taking the prepotential expression of the SM field and reinterpreting it as the gauge transformation of what will turn out to be the vector multiplet (VM), describing the supersymmetric generalization of the form with $p=1$ in table (\ref{E:p}). 

Thus, the vector multiplet prepotential is a real scalar superfield ${V}$ with the abelian gauge transformation 
\begin{align}
\delta {V} = d^2 \lambda ,
\end{align}
given in terms of the scalar multiplet of the previous subsection. 
Due to the identity (\ref{dred}), we can define the (abelian) vector multiplet field strength $W_\alpha$ as
\begin{align}
\label{E:VMconstraint}
W_\alpha := d^2 d_\alpha {V} .
\end{align}
This solves the Bianchi identity
\begin{align}
\label{VMBI}
d_{\langle \alpha} W_{\beta\rangle} = 0 
~~~\Leftrightarrow~~~
d\gamma_a W = 0 ,
\end{align}
as follows from (\ref{dred}) and the second equation in (\ref{dalgebra}).

Then the Maxwell action 
\begin{align}
S = \frac1{2} \int d^{4} x \int d^4 \theta \, {V} d^\alpha d^2 d_\alpha {V}
\end{align}
is invariant.
The equation of motion following from this action is 
\begin{align}
d^\alpha W_\alpha {{}={}} 0.
\end{align}
Together with (\ref{VMBI}), this gives the Dirac equation for $W$, as follows: The first equation of (\ref{VMBI}), when taken together with the equation of motion implies $d_{[\alpha } W_{\beta]} {{}={}} 0$. This gives $d^2W_\alpha = - p_{\alpha\beta} W^\beta$ and $\Delta_{\alpha \beta}W^\beta = - \tfrac54 p_{\alpha \beta} W^\beta$. Together, they imply the Dirac equation $0{{}={}} d_\alpha d_\beta W^{\beta} = \tfrac14 p_{\alpha \beta} W^\beta$. 

The $8+8$ components of the real scalar prepotential ${V}$ are given by
\begin{align}
\label{V}
{V} \sim {V} + \theta\eta + \theta^2f + \theta \gamma^a\theta A_a + \theta^2 \theta \chi + \theta^4 D.
\end{align}
Under a gauge transformation with SM gauge parameter
\begin{align}
\label{VMgaugeParameter}
d^2 \lambda \sim \Lambda + \theta \psi  + \theta \gamma^a \theta \partial_{ab}\lambda^b+ \theta^2 D_\lambda
\end{align}
we see that there is a Wess-Zumino gauge 
\begin{align}
\label{WZ}
{V} \sim  \theta \gamma^a\theta A_a + \theta^2 \theta \chi + \theta^4 D
~~~\mathrm{with}~~~ 
\delta A_a = \partial_{ab} \lambda^b.
\end{align}
The components of the field strength are
\begin{align}
W_\alpha \sim \chi_\alpha +(\theta\gamma^{ab})_\alpha F_{ab} + \theta_\alpha D
~~~\mathrm{with}~~~ 
F_{ab} = \tfrac12 \epsilon_{ab}{}^{cde} \partial_{cd} A_e.
\end{align}
The 2-form component satisfies the identity $\partial^{ab}F_{bc}=0$. In table \ref{E:p} it is represented by its Hodge dual $F_{abc}$ so that, indeed, this is the 1-form (as opposed to the $\ast 1$-form). 
This gives an off-shell $4+4$ component description that reduces to $2+2$ components on-shell.

\subsubsection{Gauge-invariant TM/VM Interactions}
\label{S:Interactions}
We are unable to give the VM gauge-invariant self-interactions. Nevertheless, we find that we are able to construct interesting gauge-invariant terms that are not the kinetic action. We modify the usual argument to include a term $d^\alpha W_\alpha + f({V}){{}={}} 0$ with dependence on ${V}$ instead of $W$. Then $0{{}={}} d^2 f = d^2 {V} f^\prime + 2d^\alpha {V} d_\alpha {V} f^{\prime \prime}$ implies that 
\begin{align}
\label{E:VMinteractions}
f({V}) = c_{FI} + g(\varphi)
\end{align}
where $c_{FI}$ is a constant and $\varphi^i$ are other fields that are not ${V}$. Since $f$ is independent of ${V}$, these terms integrate to $\int d^{4}x d^4\theta {V} f(\varphi)$. Now, gauge invariance requires $d^2f=0$ {\em off-shell} so
\begin{align}
\label{E:NLTM}
g_i d^2 \varphi^i  + 2g_{ij} d^\alpha \varphi^i d_\alpha \varphi^j = 0
\end{align}
where $g_i := \partial g/\partial \varphi^i$ and $g_{ij} := \partial^2 g/\partial \varphi^i\partial \varphi^j$. Thus, assuming only linear constraints on $\varphi$, we find that $g_{ij} = 0$ and either $g_i=0$ or $\varphi = {G}$ is a tensor multiplet. We now consider these terms in more detail.

\paragraph{Fayet-Iliopoulos Term} 
Although we are unable to construct minimal coupling terms and self-interactions, there is the possibility of a Fayet-Iliopoulos term
\begin{align}
S_{FI} =  {c_{FI}} \int d^{4} x \int d^4 \theta \, {V}.
\end{align}
This term is gauge invariant and supersymmetric by the usual arguments.

\paragraph{\boldmath $BF$ Coupling and Tensor Multiplet Mass} 
Consider the coupling
\begin{align}
S_{BF} = {c_{BF}} \int d^{4} x \int d^4 \theta \, {V} {G},
\end{align}
to a tensor multiplet. This is gauge-invariant due to the constraint on ${G}$. The Fayet-Iliopoulos term results if we give a vev $\vev{{G}} = {c_{FI}}/{c_{BF}}$. By solving ${G} = d^\alpha d^2 \eta_\alpha$ for its pre-potential, we can also recast the action in the form
\begin{align}
S_{BF} = {c_{BF}} \int d^{4} x \int d^4 \theta \, W^\alpha \eta_\alpha,
\end{align}

The bosonic terms in the component action are of the form ${c_{BF}} \int d^{4} x ( \epsilon^{abcde}A_a \partial_{bc} A_{de} + D{G})$. The first term can be rewritten as $B\wedge F$ where $B_{ab} \sim A_{ab}$ is the gauge 2-form in the TM and $F_{abc} \sim \partial_{[ab} A_{c]}$ is the 1-form field strength. The second term, when combined with the Maxwell action, gives rise to a mass $m=c_{BF}$ for the scalar in the TM ${G}$. 

\subsection{Dual Vector Multiplet}
\label{S:dVM}
We momentarily digress from the de Rham progression TM $\to$ SM $\to$ VM to work out the dual vector multiplet (dVM) obtained by interchanging the VM constraint with its equation of motion. In doing so, we will find the supersymmetric generalization of the form with $p=\ast1$ in table \ref{E:*p}.
The dual vector multiplet $\W_\alpha$ is defined by the constraint
\begin{align}
d^\alpha \W_\alpha = 0
~~~\Rightarrow~~~
\W_\alpha = d^2 d^\beta h_{\alpha \beta}
~:~
\delta h^c = d^2 \lambda^c
\end{align}
where the prepotential and gauge parameter satisfy $X_{\alpha \beta} = \tfrac12X_{\vev{\alpha \beta}}$ for $X=h, \lambda$.
The dual Maxwell action is
\begin{align}
S = -\frac1{2} \int d^{4} x \int d^4 \theta \,  d_\alpha h^{\alpha \gamma}  d^2 d^\beta h_{\beta\gamma},
\end{align}
giving the equation of motion 
\begin{align}
\label{E:dVMEoM}
d_{\langle \alpha} \W_{\beta \rangle} {{}={}} 0 
~~~&\Leftrightarrow~~~
d\gamma_a \W {{}={}} 0 . 
\end{align}
The components of the gauge parameter are those of a scalar multiplet (\ref{VMgaugeParameter}) with a vector index tacked on
\begin{align}
d^2 \lambda^c \sim \Lambda^c + \theta \psi^c  + \theta \gamma^a \theta \partial_{ab}\lambda^{bc}+ \theta^2 L^c .
\end{align}
This can be used to put the prepotential into the Wess-Zumino gauge
\begin{align}
\label{E:FixThis}
h^c & \sim h^c + \theta \hat \chi^c + \theta^2 a^c 
	+ \theta \gamma^a \theta A_a^c 
	+ \theta^3\left( \gamma^c \tilde \lambda + \psi^c\right) + \theta^4 D^c \cr
	& \to \theta \gamma^a \theta A_a^c 
	+ \theta^3\left( \gamma^c \tilde \lambda + \psi^c\right) + \theta^4 D^c,
\end{align}
with the residual gauge transformation $\delta A^c_a = \partial_{ab}\lambda^{bc}$, identifying this multiplet as the $p=\ast 1$ form (cf. table \ref{E:*p}).

The remaining fields enter the field strength superfield as
\begin{align}
\W_\alpha \sim {}^\ast\lambda_\alpha 
	+ (\gamma^{ab}\theta)_\alpha  {}^\ast F_{ab} + (\gamma^{a}\theta)_\alpha {}^\ast F_a
\end{align}
where $ {}^\ast F_{ab} \sim \partial_{[ca}A^c_{b]}$ and ${}^\ast F_a \sim \epsilon_a{}^{bcde}\partial_{bc}A_{de}$ are the field strength for a dual vector ({\it i.e.}~a $\ast 1$-form) and the (dual of the) field strength for a 2-form, respectively. The latter is auxiliary, being set to zero on-shell by the equation of motion (\ref{E:dVMEoM}).

\subsubsection{VM \texorpdfstring{$\leftrightarrow$}{\textleftarrow\textrightarrow} dVM Duality} 
This multiplet can be obtained by performing a duality transformation on the vector multiplet of the previous section. The duality functional for 1-form $\leftrightarrow \ast 1$-form is 
\begin{align}
\int d^{4} x \int d^4 \theta \, \left(A_\alpha d^2 d^\alpha {V} - \tfrac12 A_\alpha d^2 A^\alpha\right) .
\end{align}
Indeed, integrating out $A$ gives $A_\alpha = d_\alpha {V}$ (up to terms annihilated by $d^2$ {\it e.g.}~$d_\alpha d^2k$). Substituting back into the action gives the vector multiplet of section \ref{S:VM}. Alternatively, integrating out ${V}$ gives the constraint $d^\alpha d^2 A_\alpha =0 $ which is satisfied by the dual vector multiplet solution $A_\alpha = d^\beta h_{\alpha \beta}$ (up to terms in the kernel of $d^2$).
Again, there is a dual action,
\begin{align}
Çd^4xÊd^4ϼ\left( A^Œ d^2 d^º h_{Œº} - \f12 A^{Œ}d^2 A_{Œ} \right) .
\end{align}

\subsubsection{No Gauge-invariant dVM Interactions}
The equation of motion $d\gamma_a \W {{}={}} 0$ is annihilated by $d^2$. We can repeat the analysis of section \ref{S:Interactions} with the difference that now $f_a(h)$ is a 1-form. The condition that $d^2 f_a {{}={}} 0$ is strengthened by gauge invariance to 
conservation $d^2f_a = 0$ off-shell. Analogously to (\ref{E:VMinteractions}), we find that $f_a$ is a linear function of TMs. But then it is easy to see that the only such function consistent with Lorentz invariance is $f_a \equiv 0$. We conclude that the dVM does not admit any interactions.

\section{Supergravity}
\label{S:Fgravity}
In this section, we return to construction of the de Rham complex by extending equation (\ref{E:dR}) to the right: 
\begin{align}
\dots \longrightarrow 
	\mathrm{TM} \longrightarrow 
	\mathrm{SM} \longrightarrow 
	\mathrm{VM}\longrightarrow 
	\mathrm{sConf}\longrightarrow
	\dots
\end{align}
This means that we have a prepotential with transformation of the form of the defining constraint (\ref{E:VMconstraint}) of the VM. This is recognizable as the linearized conformal supergravity prepotential transformation. Surprisingly, then, we do not generate the supersymmetric version of a $p$-form as one might have expected.\footnote{This is also true in ordinary 5D, $N=1$ \cite{Gates:2014cqa} and 6D, $N=(1,0)$ \cite{Arias:2014ona} versions of the superspace de Rham complex and may, in fact, hold generally \cite{Linch:2014iza}. The case of 4D, $N=1$ (and lower) is special because of the possibility of producing an irreducible superfield using chirality: The 1-form constraints are $d^\alpha W_\alpha - \bar d_{\dot \alpha} ÑW^{\dot \alpha} = 0$ and $\bar d_{\dot \alpha} W_\alpha - d_\alpha ÑW_{\dot \alpha} =0$ (cf. eq. \ref{E:VMreduction}) but the latter reduces to the chirality condition (cf. \S \ref{S:F2M}). Had this not been the case, we would have found $\delta {H} _a = \bar d \tilde \sigma_a L -  d \sigma_a \bar L$ (the conformal graviton) as the next multiplet in the list. Instead, we impose chirality and extend the complex using the other identity. By doing so, we successfully generate the gauge 3-form multiplet. (See also ref. \cite{Linch:2014iza}.)
} 

\subsection{Local Superconformal Transformations}
\label{S:Superconformal}
The local superconformal transformation of the supergravity prepotential is 
\begin{align}
\label{E:FDiff}
\delta {H}_a = 2 d \gamma_a L .
\end{align}
The expansion for the gauge parameter is
\begin{align}
L_\alpha \sim L_\alpha 
	&+ \theta_\alpha \xi + (\theta \gamma^a)_\alpha \xi_a + (\theta \gamma^{ab})_\alpha \xi_{ab}
	+ \theta^2 \varepsilon_\alpha + \theta\gamma^a \theta \hat \Lambda_{a \alpha}  \cr
	&+ \theta^3_\alpha \sigma + (\theta^3 \gamma^a)_\alpha \sigma_a + (\theta^3 \gamma^{ab})_\alpha \omega_{ab}
	+ \theta^4 \eta_\alpha .
\end{align}
As we explain momentarily (c.f. \S \ref{S:FconstraintOrigins}), there is a gauge transformation of this gauge parameter that allows us to remove the first two components. In this Wess-Zumino gauge, the gauge transformation of ${H}$ is
\begin{align}
d\gamma_aL &\sim \xi_a 
	+ \theta \left( \hat \Lambda_a + \gamma_a \varepsilon 
	\right)
	+ \theta^2\left( \sigma_a  + \partial_{ab} \xi^b\right)
	+ \theta\gamma_a \theta \left( \sigma  +\partial_{bc} \xi^{bc}\right)
\cr
&+ \theta\gamma^{b}\theta \left(\omega_{ab} 
	+ \partial_{a}{}^c \xi_{bc}\right)
	+ \theta^3\left( \gamma_a \eta + \partial_{ab} \gamma^b \varepsilon +\partial_{ab} \hat \Lambda^b \right)
	+ \theta^4 \partial_{ab} \sigma^b
\end{align}
The component fields parameterize:
\vskip12pt
\hspace{12mm}
{\renewcommand{\arraystretch}{1.5} 
\begin{tabular}{|c|c|c|c|c|c|}
 \hline
parameter 
	&$\xi_{ab}$
		&$\varepsilon^\alpha$
		&$\sigma$ 
		&$\omega_{ab}$
		&$\eta_\alpha$ 
\\ \hline	
local type 
	&translation
		& supersymmetry
		&scale
		&Lorentz
		&$S$-supersymmetry
\\  \hline		
\end{tabular}
}
\vskip12pt

\noindent
The parameters $\{\xi_a, \hat \Lambda_{a\alpha}, \sigma_a\}$, together with local scale (for $\varphi$), local Lorentz (for $b_{ab}$), and local $S$-supersymmetry (for the trace $\psi$ of the gravitino), can be used to reduce the components of the supergravity prepotential to the Wess-Zumino gauge\footnote{The component $\hat \Lambda_{a \alpha}$ is Lorentz reducible: It contains a $\gamma$-trace that cannot be absorbed by redefining $\varepsilon_\alpha$ (because $\theta \gamma^a \theta (\gamma_a \Lambda)_\alpha \sim \theta^2\Lambda_\alpha + \theta_\alpha (\theta\Lambda)$ and a redefinition of $\varepsilon$ can only shift away the first term). Similarly, the $\hat \chi_{a\alpha}$ component is reducible and we use all of $\hat \Lambda$ to gauge away all of $\hat \chi$.
} 
\begin{align}
{H}_a & \sim {H}_a + \theta \hat \chi_a + \theta^2 a_a 
	+ \theta \gamma^b \theta \left( \eta_{ab} \varphi + h_{ab} + b_{ab}\right)
	+ \theta^3\left( \gamma_a \psi + \psi_a\right) + \theta^4 A_a \cr
	& \to \theta \gamma^b \theta h_{ab} 
	+ \theta^3\psi_a + \theta^4 A_a .
\end{align}
The remaining gauge transformations
\begin{align}
\delta h_{ab} = \partial_{(a}{}^{c} \xi_{b)c} 
,~~
\delta \psi_{a \alpha} = \partial_{ab} \gamma^b \varepsilon 
	-\mathrm{trace}
,~\mathrm{and}~~
\delta A_a = 0, 
\end{align}
represent the {\bf F}-theory analogues of the usual infinitesimal diffeomorphism and local supersymmetry transformations.

\subsubsection{Geometrical Origin of the F-constraint} 
\label{S:FconstraintOrigins}
The gauge transformation $\delta L_\alpha$ leaving $\delta {H}_a$ invariant (gauge for gauge) corresponds to the vector multiplet field strength of section \ref{S:VM}: 
$$ 0 = d_{\langle \alpha} \delta L_{\beta\rangle}â\Rightarrowâ\delta L_\alpha = d^2 d_\alpha C $$
for an arbitrary real scalar field $C$ provided (cf.~eq.~\ref{E:d4b})
$$ 0 = d_{\langle \alpha} d^2 d_{\beta\rangle} = 2p^2_{\vev{\alpha \beta}}. $$
This gives the manifestly supersymmetric generalization of the observation in the Introduction that the F-constraint follows from the gauge invariance of the gauge invariance, as is also true in {\bf T}-theory.  Also as in {\bf T}, at the nonlinear level it follows from the closure of the algebra of generalized diffeomorphisms, as was shown in bosonic {\bf F}-theory \cite{Berman:2011cg}.

\subsection{Action}
\label{S:SGAction}
Given the {\bf F}-theory action $S_{\bf F}$, the equation of motion is defined by
\begin{align}
G_a := { \delta S_{\bf F}\over \delta {H}^a } {{}={}} 0 .
\end{align}
Due to the gauge transformation of (\ref{E:FDiff}), this superfield must satisfy
\begin{align}
\label{E:EinsteinTensor}
\delta G_a = 0
~~~\Leftrightarrow~~~
d^\alpha G_{\alpha \beta} = 0.
\end{align}
Since we are working in the quadratic approximation, the action is determined to be
$S_{\bf F} = {1\over 2} \int d^{4}x \int d^4\theta \, {H}^a G_a({H})
$ 
once we have found such a $G$. We now turn to the construction of the linearized Einstein tensor.

Let\footnote{Note that $\Delta_{\langle \alpha}{}^\gamma H_{\beta\rangle \gamma} \equiv 0$ (in vector notation, for example, there is no way to make a non-zero vector out of $\Delta_a H_b$). Therefore, 
	$0=H^{\alpha \beta} d^2 \Delta_{\alpha}{}^\gamma H_{\beta\gamma} 
		= H^{\alpha \beta} p_{\langle \gamma}{}^\delta \Delta_{\alpha\rangle \delta}H_\beta{}^\gamma 
		=H^{\alpha \beta} (p^{\gamma\delta}\Delta_{\delta \alpha}H_{\beta \gamma} 
		+\Delta^{\gamma\delta}p_{\delta \alpha}H_{\beta \gamma})$.
} 
\begin{align}
\label{E:SuperEinsteinF}
G_{\alpha \beta} = 
	a_0  d^4 {H}_{\alpha \beta} 
	+ a_1  d^2 p_{\langle \alpha}{}^\gamma {H}_{\beta\rangle \gamma} 
	+ {a_2} \Delta^{\gamma \delta} p_{\gamma \langle \alpha} {H}_{\beta\rangle \delta} 
	+ a_3 p_{\langle \alpha}{}^\gamma p_{\beta \rangle}{}^\delta {H}_{\gamma \delta} 
	+ a_4 p^2 {H}_{\alpha \beta}
\end{align}
Although we can fix the coefficients by demanding invariance under the gauge transformation (\ref{E:FDiff}), it turns out to be simpler to impose the Bianchi identity. Contracting $d^\beta$ on the first three terms gives
\begin{align}
d^\beta d^4 {H}_{\alpha \beta} 
	&= -p_\gamma^\beta d^{3\gamma} H_{\alpha \beta} 
\cr
d^\beta d^2 p_{\langle \alpha}{}^\gamma {H}_{\beta\rangle \gamma} 
	&= -p_\gamma^\beta d^{3\gamma} H_{\alpha \beta} 
		-p_\alpha^\beta d^{3\gamma} H_{\beta\gamma} 
		+\tfrac12 p_{\langle \alpha}^\gamma p_{\beta\rangle}^\delta d^\beta H_{\gamma \delta}
		-\tfrac12 p^2 d^\beta H_{\alpha \beta}
\cr
d^\beta \Delta^{\gamma \delta} p_{\gamma \langle \alpha} \delta {H}_{\beta\rangle \delta } 
	&= p_\gamma^\beta d^{3\gamma} H_{\alpha \beta} 
		-p_\alpha^\beta d^{3\gamma} H_{\beta\gamma} 
		-\tfrac12 p_{\langle \alpha}^\gamma p_{\beta\rangle}^\delta d^\beta H_{\gamma \delta}
		-\tfrac12 p^2 d^\beta H_{\alpha \beta}
\end{align}
Therefore, $d^\beta G_{\alpha \beta}=0$ implies $(a_1, a_2, a_3, a_4) = (-1, 1, 1, 0) \tfrac{a_0}2$, and the action 
\begin{align}
\label{E:Faction}
S_{\bf F} 
	&= {a_0\over 4} \int d^{4}x \int d^4\theta \, 
	\Big\{
		H^{\alpha \beta} d^4 H_{\alpha \beta} 
		+ p_\alpha{}^\gamma H^{\alpha \beta} d^2 H_{\beta \gamma} 
		+ p_\alpha{}^\gamma H^{\alpha \beta} p_\beta{}^\delta H_{\delta\gamma}
		- p_{\gamma\alpha} H^{\alpha \beta} \Delta^{\gamma \delta} H_{\beta \delta}
\Big\}
\end{align}
is invariant.

Alternative forms of this action can be derived using the identities in appendix \ref{S:Sp4}. For example, by solving for the $Hp\Delta H$ term in the sum of the last two equations in (\ref{dalgebra}), 
the action can be put in the form
\begin{align}
S_{\bf F} 
	&= {3 a_0\over 8} \int d^{4}x \int d^4\theta \, 
	\Big\{
		H^{\alpha \beta} d^4 H_{\alpha \beta} 
		+\tfrac13 p_\alpha{}^\gamma H^{\alpha \beta} d^2 H_{\beta \gamma} 
		+ p_\alpha{}^\gamma H^{\alpha \beta} p_\beta{}^\delta H_{\delta\gamma}
		+ \tfrac16 (\Delta_{\alpha \beta} H^{\alpha \beta})^2
	\Big\} .
\end{align}
This form can be reduced further to one with only $p$ and $\Delta$ terms giving
\begin{align}
S_{\bf F} 
	&= -{a_0\over 8} \int d^{4}x \int d^4\theta \, 
	\Big\{
		\Delta_{\alpha \beta} H^{\gamma \delta} \Delta_{\gamma \delta} H^{\alpha \beta}  
		+2p_{\gamma \alpha} H^{\alpha \beta} \Delta^{\gamma \delta} H_{\delta \beta } 
\Big\} .	
\end{align}
In section \ref{S:MSG}, we will reduce the {\bf F} action in the form (\ref{E:Faction}) to {\bf M}. Matching to the quadratic action of old-minimal supergravity fixes the overall normalization to the value $a_0 = -\tfrac13$.

\subsubsection{No Weyl tensor for {\bf F}}
\label{S:NoWeyl}
The linearized Einstein tensor $G_a$ is not the field strength of the conformal graviton ${H}_a$ since it vanishes on-shell. Next, we explain why we are unable to construct the linearized super-Weyl tensor $W_{\alpha \beta \gamma}$ of dimension $\tfrac52$ invariant under the gauge transformation (\ref{E:FDiff}). The linearly independent terms of dimension $\tfrac32$ and their transformations are
\begin{align}
\delta p_{(\alpha}{}^\delta d_\beta {H}_{\gamma)\delta} &= 
	-\tfrac12 p_{(\alpha \beta} p_{\gamma)\delta}L^\delta 
	+\tfrac12 p_{(\alpha \beta}\Delta_{\gamma)\delta}L^\delta 
	-\tfrac14 p_{(\alpha \beta} d^2 L_{\gamma)} 
	+\tfrac 12 d_{(\alpha}d^2 d_\beta L_{\gamma)} 
\cr %
\delta p_{(\alpha \beta}d^\delta {H}_{\gamma)\delta} &= 
	-\tfrac12 p_{(\alpha \beta} p_{\gamma)\delta}L^\delta 
	+\tfrac32 p_{(\alpha \beta}\Delta_{\gamma)\delta}L^\delta 
	-\tfrac54 p_{(\alpha \beta} d^2 L_{\gamma)} \cr
\delta \Delta_{(\alpha}{}^\delta d_\beta {H}_{\gamma)\delta} &= 
	-p_{(\alpha \beta}\Delta_{\gamma)\delta}L^\delta 
	-\tfrac32 p_{(\alpha \beta} d^2 L_{\gamma)} 
	+\tfrac 12 d_{(\alpha}d^2 d_\beta L_{\gamma)} 
\end{align}
Note that every term on the right-hand-side either contains $p_{(\alpha \beta)}$ or is of the form $d_{(\alpha}d^2 d_\beta L_{\gamma)}= 2\Delta_{\delta (\alpha}p_\beta^\delta L_{\gamma)}$. The former vanish when reduced to {\bf M} and restricted to undotted spin indices whereas the latter are annihilated by $d^2$. A related observation is that $d^2d_{(\alpha} p_\beta{}^\delta  h_{\gamma)\delta} \to W_{\alpha \beta \gamma}$, by itself, reduces to the {\bf M}-theory Weyl tensor when keeping only holomorphic spinor indices.\footnote{The same was true of the VM superfield strength $W_\alpha$ when it was written in terms of the prepotential: There too the holomorphic restriction of {\bf F} $\to$ {\bf M} automatically spits out the {\em chiral} superfield $W_\alpha = \bar {d}^2{d}_\alpha V$.} 

With this, it is clear that we can find a linear combination canceling the terms of the form $p_{(\alpha \beta} p_{\gamma)\delta}L^\delta$ and $p_{(\alpha \beta} d^2 L_{\gamma)} $. The terms of the form $d_{(\alpha}d^2 d_\beta L_{\gamma)}$ are canceled once we hit the whole thing with $d^2$. This leaves the terms $ d^2 p_{(\alpha \beta} \Delta_{\gamma)\delta}L^\delta$ with the coefficient $-1-\tfrac23\neq 0$.
Therefore, we conclude that it is impossible to construct a Weyl tensor superfield in {\bf F} theory.
 
In retrospect, this can be explained by a similar observation in {\bf T}. There, it is known that one cannot construct the T-invariant Weyl tensor \cite{Siegel:1993th}. Since {\bf F} $\Rightarrow$ {\bf T}, we conclude that ``no {\bf T} Weyl tensor'' $\Rightarrow$ ``no {\bf F} Weyl tensor''.

\section{Reduction F \texorpdfstring{$\to$}{\textrightarrow} M, T, S}
\label{S:Reduction}

The signature of our ten-dimensional toy {\bf F}-theory is $(4,6)$ with 4 space-like and 6 time-like directions. In particular, in flat space
\begin{align}
\Box := - \eta^{AB} \partial_A \partial_B := -\tfrac12 \partial_{ab}\partial^{ab} = 
	-\partial_{-10}^2 + \partial_{-1i}^2 - \tfrac12 \partial_{ij}\partial^{ij}  + \partial_{0i}^2 
\end{align}
with the isometries $SO(4,6)$. (Here the ${}_{-1}$ index is the timelike one of \S \ref{S:Review}
and we chose the sign so that $\Box \to \Box$ under {\bf F}$\to${\bf M}.)
Reduction to {\bf M}, {\bf T}, and {\bf S} proceeds by solving the F-constraint $\partial_{[ab} \partial_{cd]}=0$ (\ref{E:Fconstraint}) and dimensional reduction in various combinations, as explained in the Introduction.

\subsection{F \texorpdfstring{$\to$}{\textrightarrow} M} 
\label{S:F2M}
Reducing to {\bf M} is achieved solving the F-constraint in the hyperplane orthogonal to time-like vector $u$. Fixing $u^a = \delta^a_{-1}$, this means that we are keeping only those $p_{ab}$ with one ${}_{-1}$ index and, therefore, $\Box \to - \partial_{-10}^2 + \partial_{-1i}^2 $ reduces to the Laplacian in signature $(3,1)$. 

The superspace derivatives reduce to $d_\alpha, \bar d_{\dot \alpha}$ with the familiar form for the four-dimensional, $N=1$ supersymmetry algebra (our conventions are explained in detail in appendix \ref{S:FtoM})
\begin{align} 
\{ d, d\} = 0 
,~~~
\{ d_\alpha , \bar d_{\dot \alpha} \} = p_{\alpha \dot \alpha}
,~~~
\{ \bar d, \bar d\} = 0 .
\end{align}
The superspace theories of section \ref{S:Matter} reduce to combinations of four-dimensional representations and the reduction of the gravity theory turns out to be a modified version of old minimal supergravity that has a real prepotential (3-form multiplet) for its chiral compensator. We now turn to the derivation of these results (they are summarized in table \ref{T:M2}).

\subsubsection{Chirality} 
\label{S:MChirality}
{\bf M}-theory has chiral fields. We now explain how these emerge from a manifestly real {\bf F}-theory. Consider again the vector multiplet field strength of section \ref{S:VM}. We reduce the constraint (\ref{VMBI}) 
\begin{align}
\label{E:VMreduction}
d_{\langle \alpha} W_{\beta\rangle} = 0 \longrightarrow 
\begin{cases}
{d}^\alpha W_\alpha - \bar {d}^{\dot \alpha} ÑW_{\dot \alpha} = 0\\
\bar {d}_{\dot \alpha} W_\alpha - {d}_\alpha ÑW_{\dot \alpha} = 0
\end{cases}
\end{align}
(see also appendix \ref{S:FtoM}).
The second equation is weaker than the chirality condition $\bar {d} W= 0$ but it implies $p_{(\alpha}^{(\dot \alpha} \bar {d}^{\dot \beta)} W_{\beta)} = 0$. This suffices to conclude that $\bar {d}_{\dot \alpha} W_\alpha = $ constant. Thus, chirality emerges as a choice of boundary condition in the space of {\bf F}-fields.

Alternatively, we can solve the constraint in terms of the prepotential $W_\alpha = d^2 d_\alpha V$. Upon reduction, $W_\alpha \to W_\alpha, \overline W_{\dot \alpha}$, the new $W=\bar d^2d_\alpha V$ is automatically chiral because $d^3\equiv 0$. 

\subsubsection{{\bf M} Fields}
The reduction of the fields from {\bf F} $\to$ {\bf M} is subtle due to the tension between the reality of the former and the chirality of the latter. In this section, we will demonstrate the matching of representations by comparing them to table \ref{T:M} of four-dimensional $p$-forms \cite{Gates:1980ay,Gates:1983nr}:

\begin{table}[h]
$$\vcenter{\tabskip=0pt \offinterlineskip
\halign{\strut#& \vrule#\tabskip=1em plus2em& \hfil$#$\hfil& \vrule#& \hfil$#$\hfil& \vrule#& \hfil$#$\hfil& \vrule#& \hfil#\hfil& \vrule#& \hfil$#$\hfil& \vrule#\tabskip=0pt\cr
	\noalign{\hrule}
        \vrule height13pt depth3.5pt width0pt && p¼&& ßA_p && \widehat{dA}_p && multiplet && \hbox{EOM} &\cr
	\noalign{\hrule}
        \vrule height13pt depth3.5pt width0pt && 0 && \Phi && i(\Phi-\bar \Phi) && scalar && \bar d^2\widehat{dA} &\cr
        \vrule height13pt depth3.5pt width0pt && 1 && V && \bar d^2 d_Œ V && vector &&  d^Œ\widehat{dA}_Œ + \mathrm{h.c.}&\cr
        \vrule height13pt depth3.5pt width0pt && 2 && \Phi_Π&& d^\alpha \Phi_\alpha+ \mathrm{h.c.} && tensor && d_\alpha \widehat{dA} &\cr
        \vrule height13pt depth3.5pt width0pt && 3 && V && \bar d^2 V && gauge 3-form && d^2\widehat{dA} + \mathrm{h.c.}&\cr
        \vrule height13pt depth3.5pt width0pt && 4 && \Phi && \bar d_{\dot \alpha}\Phi \equiv 0 && gauge 4-form && 0 &\cr
	\noalign{\hrule}}}$$
\caption{4D, $N=1$ $p$-forms: 
$(\delta \widehat A_p, \widehat F_{p+1}, \widehat {dF}_{p+1}) = (\widehat {d\Lambda}_{p-1}, \widehat {dA}_{p}, 0)$.
$\Phi$s are chiral and $V$s are real.}
\label{T:M}
\end{table}

\paragraph{SM $\to$ 0-form and 3-form} 
As alluded to in section \ref{S:SM}, the scalar multiplet reduces as $\phi = d^2 U \to d^2U + \bar d^2U= (\bar \Pi +\Pi)$ to a chiral field $\Pi : = \bar d^2 U$ with real prepotential. Comparing to the table, we find that this is both of the form of the 0-form field strength {\em and} that of the gauge 3-form multiplet. The 4-form field strength of the 3-form $C_{abc}\sim \epsilon_{abcd}[d,\bar d]^{d}u$ is the lowest component of the combination $d^2\Pi -\bar d^2 \bar \Pi \sim \epsilon^{abcd}\partial_aC_{bcd}$. The gauge transformation $\delta U = d^\alpha d^2 \lambda_\alpha$ implies that $\delta C_{abc} = \partial_{[a} \lambda_{bc]}$, as we now show.

\paragraph{TM $\to$ 2-form} 
The tensor multiplet has field strength $G= d^\alpha d^2 \eta_\beta$ reducing to $G= d^\alpha \bar d^2 \eta_\alpha + \bar d^{\dot \alpha} d^2 \bar \eta_{\dot \alpha}$ so we recover the 3-form field strength of the four-dimensional 2-form gauge potential $B_{ab}$. To conclude that this is, in fact, the gauge 2-form however, we must verify that the gauge transformation is $\delta B_{ab} = \partial_{[a} \lambda_{b]}$. The {\bf F}-theory transformation $\delta \eta_\alpha = d^2d^\beta \lambda_{\alpha \beta}$ reduces to $\delta \eta_\alpha = \bar d^2d_\alpha \lambda + d^2\bar d^{\dot \alpha} \lambda_{\alpha \dot \alpha}$ after decomposing $\lambda_{\vev{\alpha \beta}} \to (\lambda, \lambda_{\alpha \dot \alpha})$ into four-dimensional representations. The first term is the field strength of a 1-form and, thus, gives the correct transformation of the 2-form $B_{\alpha \beta}\sim d_{(\alpha} \bar d^2 \eta_{\beta)}$ because the second term cannot contribute. Since $\eta_\alpha$ is complex, the combination $\Phi_\alpha = \bar d^2\eta_\alpha$ is an arbitrary chiral field. In the description in terms of $\Phi_\alpha$ the $\lambda_{\alpha \dot \alpha}$ parameter drops out completely and we conclude that {\bf F} TM $\to$ {\bf M} TM, as expected.

\paragraph{VM $\to$ 1-form and 4-form} 
The vector multiplet field strength $W_\alpha = d^2d_\alpha V$ reduces trivially to that of the four-dimensional VM $W_\alpha = \bar d^2 d_\alpha V$ and its conjugate but, again, we should check the gauge transformation rule. This gives $\delta V \to d^2\lambda +\bar d^2\lambda$ the field strength of a 3-form rather than a chiral scalar multiplet. Contrary to the case of the tensor, this distinction persists in the description of this theory in terms of the combination $A_\alpha = d_\alpha V$, implying instead that the $(d^2-\bar d^2)V \sim \epsilon^{abcd}C_{abcd}$ with the gauge transformation $\delta C_{abcd} = \partial_{[a}\lambda_{bcd]}$. This component gauge 4-form carries no degrees of freedom on- or off-shell.

\paragraph{dVM $\to$ Dual 1-form} 
The dual vector multiplet field strength ${}^\ast W_\alpha = d^2d^\beta V_{\vev{\alpha \beta}}$ reduces into two parts ${}^\ast W_{\alpha} = i\bar d^2 d_\alpha V + d^2 \bar d^{\dot \alpha} V_{\alpha \dot \alpha}$. The first part $\widetilde W_\alpha = i \bar d^2 d_\alpha V$ gives a dual vector in four dimensions but this time we cannot remove $\Upsilon_{\dot \alpha} := \bar d^2d^\alpha V_{\alpha \dot \alpha}$. On the other hand, $d{}^\ast W = i d \widetilde W$ and $\bar d {}^\ast W = \bar d \bar \Upsilon$ so the action reduces to 
\begin{align}
\int d^4x \int d^2\theta \left( \widetilde W_\alpha \widetilde W^\alpha + \Upsilon_{\dot \alpha} \Upsilon^{\dot \alpha} \right)+ \mathrm{h.c.}
\end{align}
and the dynamics of $V$ and $V_{\alpha \dot \alpha}$ decouple.
The equation of motion for $V$ imposes $d^\alpha \widetilde W_\alpha - \mathrm{h.c.}{{}={}} 0$ on-shell and that of $V^{\alpha \dot \alpha}$ imposes $d_\alpha \Upsilon_{\dot \alpha}  - \mathrm{h.c.}{{}={}} 0$. Note that these equations are equivalent to $d^\alpha {}^\ast W_\alpha - \mathrm{h.c.}{{}={}} 0$ and $d_\alpha {}^\ast \bar W_{\dot \alpha}  - \mathrm{h.c.}{{}={}} 0$, which are dual to the four-dimensional projections of the Bianchi identity of the {\bf F}-theory vector multiplet. 
By the chirality argument in section \ref{S:MChirality}, we may assume that, on-shell, $d_\alpha \Upsilon_{\dot \alpha}{{}={}} 0$. Since $\Upsilon$ is already chiral off-shell, we conclude that it carries no degrees of freedom on-shell and we are left only with the degrees of freedom of the dual vector multiplet $\widetilde W_\alpha$. 

\begin{table}[h]
$$\vcenter{\tabskip=0pt \offinterlineskip
\halign{\strut#& \vrule#\tabskip=1em plus2em& \hfil$#$\hfil& \vrule#& \hfil$#$\hfil& \vrule#& \hfil$#$\hfil& \vrule#& \hfil#\hfil& \vrule#& \hfil$#$\hfil& \vrule#\tabskip=0pt\cr
	\noalign{\hrule}
        \vrule height13pt depth3.5pt width0pt && p¼\hbox{(mod 6)} && ßA_p && \widehat{dA}_p && multiplet && \hbox{EOM} &\cr
	\noalign{\hrule}
        \vrule height13pt depth3.5pt width0pt && 0 && U && (d^2+\bar d^2)U && scalar + 3-form && (d^2+\bar d^2)\widehat{dA} &\cr
        \vrule height13pt depth3.5pt width0pt && 1 && V && \bar d^2 d_Œ V && vector + 4-form &&  d^Œ\widehat{dA}_Œ + \mathrm{h.c.}&\cr
        \vrule height13pt depth3.5pt width0pt && 2 && L_\alpha && 
        		\left\{
        		\begin{array}{c}
        		\bar d_{\dot \alpha} L_{\alpha}+ \mathrm{h.c.}\\
			d^{\alpha} L_{\alpha}+ \mathrm{h.c.} 
			\end{array}
			\right.
			&& 
			$\begin{array}{c}
			\mathrm{superconformal}\\
			\mathrm{parameters}
			\end{array}$
			&& $none$ &\cr
        \vrule height13pt depth3.5pt width0pt && 3 && (H_a, P) && (G_a(H,P),R(H,P)) && SG && 
        \widehat{dA} &\cr
        \vrule height13pt depth3.5pt width0pt && 4 && (G_a, R) && 
		\bar d^{\dot \alpha} G_{\alpha\dot \alpha} + d_\alpha R 
        && SG Bianchi Id. && 
        \begin{array}{c}
        		\bar d_{\dot \alpha} \widehat{dA}_\alpha+ \mathrm{h.c.}\\
			d^{\alpha} \widehat{dA}_\alpha+ \mathrm{h.c.} 
			\end{array}
        &\cr
        \vrule height13pt depth3.5pt width0pt && 5 && \psi_\alpha && d^\alpha \bar d^2 \psi_\alpha + \mathrm{h.c.} && tensor && \bar d^2d_\alpha\widehat{dA} &\cr
	\noalign{\hrule}}}$$
\caption{{\bf F} $\to$ {\bf M} reduction.}
\label{T:M2}
\end{table}

\subsubsection{{\bf M} Supergravity}
\label{S:MSG}
As implied by the reduction of the spinors from {\bf F} $\to$ {\bf M} (cf. appendix \ref{S:FtoM}), the supergravity field $H_{\vev{\alpha \beta}}$ decomposes into the fields $H_{\alpha \dot \alpha}$ and $H_{-1}$ with gauge transformations (\ref{E:FDiff}) 
\begin{align}
\delta H_{\alpha \dot \alpha} = \bar d_{\dot \alpha} L_\alpha - d_\alpha \bar L_{\dot \alpha}
~~~\mathrm{and}~~~
\delta H_{-1} = d^\alpha L_\alpha -\bar d^{\dot \alpha}L_{\dot \alpha}.
\end{align}
They are recognizable as the superconformal transformations of the four-dimensional, $N=1$ supergravity prepotential $H^a$ and the chiral compensator $\chi \propto \bar d^2 H_{-1}$ of old-minimal supergravity. The fact that the prepotential $H_{-1}$ is imaginary ({\it vs}. generic complex) means that this supergravity theory has a 3-form in its chiral compensator (cf.~table \ref{T:M}) as compared to the more familiar unconstrained chiral compensator of old-minimal supergravity \cite{Grisaru:1981xm}.\footnote{This version of supergravity was found to be useful to describe supergravity-mediated supersymmetry breaking in five-dimensional brane-world scenarios in references \cite{Linch:2002wg,Buchbinder:2003qu}. 
} 

The super-Einstein tensor $G_{\vev{\alpha \beta}}$ reduces to the four-dimensional, $N=1$ super-Einstein tensor $G_{\alpha \dot \alpha}$ and the chiral superfield $R$ (containing the curvature scalar) in the combination $R+\bar R$. 
Explicitly, reducing (\ref{E:SuperEinsteinF}), we obtain 
\begin{align}
G_{\alpha \dot \alpha} 
	&= 2 a_0 d^\gamma\bar d^2 d_\gamma H_{\alpha \dot \alpha}
	-(\tfrac12 a_0 - a_3)p_{\beta \dot \beta}p^{\beta \dot \beta} H_{\alpha \dot \alpha}
	-2 a_1(d^2+\bar d^2) p_{\alpha \dot \alpha} H^\gamma{}_\gamma 
\cr& \hspace{3cm}
	+\tfrac{a_2}2 p_{(\alpha}{}^{\dot \beta} \Delta_{\beta) \dot \beta} H^\beta{}_{\dot \alpha}
	-\tfrac{a_2}2 p^\beta{}_{(\dot \beta|} \Delta_{\beta|\dot \alpha} H_\alpha{}^{\dot \beta}
	-2a_3 p_{\alpha \dot \alpha}p_{\beta \dot \beta}H^{\beta \dot \beta}		
\cr
G^\gamma{}_\gamma &= 
	2 a_0  \{d^2, \bar d^2\}  H^\gamma{}_\gamma
	+ (\tfrac12 a_0 - a_3) p_{\delta \dot \delta}p^{\delta \dot \delta} H^\gamma{}_\gamma
	+4 a_1  (d^2 +\bar d^2 ) p_{\gamma \dot \gamma}H^{\gamma \dot \gamma}
.
\end{align}
Recalling that $(a_1, a_2, a_3, a_4) = (-\tfrac12, \tfrac12, \tfrac12, 0) a_0$, the coefficients in parentheses add to zero, thereby canceling the $p^2$ terms, as required. This reduction agrees with the super-Einstein tensor $G_a$ and super-Ricci scalar $G_{-1} \propto R+\bar R$ of old-minimal supergravity \cite{Gates:1983nr} 
\begin{align}
G_{\underline a} = -\tfrac23 d^\gamma\bar d^2 d_\gamma H_{\underline a}
	+\tfrac13 p_{\underline a}p_{\underline b}H^{\underline b}
+\tfrac i6 \varepsilon_{\underline a\underline b\underline c\underline d}
			p^{\underline b} [d, \bar d]^{\underline c}H^{\underline d}
	+p_{\underline a} (\chi -\bar \chi)	
~,~~
R= \bar d^2(\bar \chi -\tfrac13p_{\underline a} H^{\underline a}),
\end{align}
provided we take $a_0=-\tfrac13$ and and the linearized chiral compensator $\chi= -\tfrac {\sqrt2}3 \bar d^2 H_{-1}$ (related to the non-linear compensator by $\varphi^3 = 1-\sqrt{2}\bar d^2 H_{-1}$). For completeness, we give the quadratic {\bf M}-theory action 
\begin{align}
\label{E:Maction}
S_{\bf M} 
	&= \int d^{4}x \int d^4\theta \, 
	\Big\{
		-\tfrac 16 H^{\underline a} d^\beta \bar d^2 d_\beta H_{\underline a}
		-\tfrac 1{12} (p_{\underline a}H^{\underline a})^2
		-\tfrac 1{36} ([d_\alpha, \bar d_{\dot \alpha}] H^{\underline a})^2
	-\tfrac13 (\chi -\bar \chi) p_{\underline a}H^{\underline a}
	+\bar \chi \chi 
	\Big\} 
\end{align}
to which the {\bf F} action (\ref{E:Faction}) reduces.

The constraint (\ref{E:EinsteinTensor}) reduces to $\bar d^{\dot \alpha} G_{\alpha \dot \alpha} = d_\alpha R$ familiar from old-minimal supergravity. 
As guaranteed by gauge invariance, the quadratic action reduces to that of linearized old-minimal supergravity with chiral compensator with a real prepotential (cf. {\it e.g.} refs. \cite{Grisaru:1981xm, Linch:2002wg}).
The {\bf M}-theory Weyl tensor $W_{\alpha \beta \gamma}$ that satisfies $\bar d_{\dot \alpha} W_{\alpha \beta \gamma}= 0$ and $d^{\alpha} W_{\alpha \beta \gamma}=\tfrac12 p_{(\alpha}{}^{\dot \gamma} G_{\beta) \dot \gamma}$ can now be constructed but, as pointed out in section \ref{S:NoWeyl}, it does not come by reduction from a superconformally invariant tensor in {\bf F}. 

\subsection{F \texorpdfstring{$\to$}{\textrightarrow} T} 
\label{S:F2T}
As explained in section \ref{S:Review}, reducing to {\bf F} $\to$ {\bf T} is achieved by truncating $p_{3a}\to 0$ (dimensional reduction). The covariant derivatives reduce to $d_\alpha \to d_\alpha, d_{\t\alpha}$ and their algebra takes the form
\begin{align} 
\{ d_\alpha , d_\beta \} = 2 p_{\alpha \beta}
,~~~
\{ d_\alpha , d_{\t \alpha} \} = 0
,~~~
\{ d_{\t \alpha} , d_{\t \beta} \} = 2 p_{{\t \alpha}{\t \beta}}.
\end{align}
The F-constraint reduces to the T-constraint
\begin{align}
\label{E:TConstraint}
p^2 - {p^{\prime 2}} = 0.
\end{align}
These constraints are invariant under Spin(2,2) $\cong$ SL(2;$\mathbf R$)$\times$SL(2;$\mathbf R$) and under the $\mathbf Z_2$ interchanging the factors, which we will call ``T-parity''. 
The reduction of the fields to {\bf T} is unfamiliar so we will work out some examples. 
Throughout this section, we reduce $\int d^4x \to \ell \int d^3x$ but then proceed to absorb $\ell$ into the definition of the fields so that no explicit factors appear. (Alternatively, we reduce and then set $\ell \to 1$.)

The basic relation in {\bf T} is 
\begin{align}
(d^2-d^{\prime 2})(d^2+d^{\prime 2}) = 0.
\end{align}
From this, it follows that a ``scalar'' multiplet $\phi = (d^2+d^{\prime 2})U$ satisfies $(d^2-d^{\prime 2})\phi=0$ but also that its prepotential has the gauge transformation $\delta U =(d^2-d^{\prime 2})\lambda$. If we define the ``vector'' multiplet prepotential $V$ to have this field as a gauge transformation $\delta V= (d^2+d^{\prime 2})\lambda$, we find that its field strength is given by the other combination $(d^2-d^{\prime 2})V$. This gives the {\bf T}-form table \ref{T:T1}.

\begin{table}[h]
$$\vcenter{\tabskip=0pt \offinterlineskip
\halign{\strut#& \vrule#\tabskip=1em plus2em& \hfil$#$\hfil& \vrule#& \hfil$#$\hfil& \vrule#& \hfil$#$\hfil& \vrule#& \hfil#\hfil& \vrule#& \hfil$#$\hfil& \vrule#\tabskip=0pt\cr
	\noalign{\hrule}
        \vrule height13pt depth3.5pt width0pt &&p$ (mod 2)$ && ßA_p && \widehat{dA}_p && multiplet && \hbox{EOM} &\cr
	\noalign{\hrule}
        \vrule height13pt depth3.5pt width0pt && 0 && U && (d^2+d^{\prime 2})U && ``scalar'' && (d^2+\bar d^2)\widehat{dA} &\cr
        \vrule height13pt depth3.5pt width0pt && 1 && V && (d^2 - d^{\prime 2}) V && ``vector'' &&  (d^2-d^{\prime 2})\widehat{dA} &\cr
	\noalign{\hrule}}}$$
\caption{Super T-forms: $(\delta \widehat A_p, \widehat F_{p+1}, \widehat {dF}_{p+1}) = (\widehat {d\Lambda}_{p-1}, \widehat {dA}_{p}, 0)$.}
\label{T:T1}
\end{table}

We now compare this to the reduction {\bf F} $\to$ {\bf T} as represented in table \ref{T:T}.
The actions for the scalar and tensor multiplets ``reduce'' simply by plugging in the explicit forms for 
$\phi = d^2 U + d^{\prime 2}U$ 
	and 
$G = d^\alpha (d^2 +d^{\prime 2})\eta_\alpha + d^{\alpha^\prime} (d^2 +d^{\prime 2})\eta_{\alpha^\prime} 	= (d^2 -d^{\prime 2})v$ with $v:=d^{\alpha^\prime} \eta_{\alpha^\prime}-d^\alpha\eta_\alpha$. The vector multiplet action $\int  d^\alpha V W_\alpha$ reduces to $\int (d^\alpha V W_\alpha+ d^{\alpha\prime} V W_{\alpha\prime}) = \int \Gamma^2$ with $\Gamma := (d^2-d^{\prime 2})V$ after integration by parts, taking the same form as the tensor multiplet theory with the replacement $v \to V$.

By contrast, the dual vector multiplet action can be written as
\begin{align}
- \int d^3x \int d^2\theta d^2\theta^\prime \, \left(
		\Gamma^2 + 2 d_\alpha d_{\alpha^\prime} v^{\alpha\alpha^\prime} \Gamma
		-2 d_\alpha d_{\alpha^\prime} v\Gamma^{\alpha\alpha^\prime} 
		-\tfrac12 \Gamma_{\alpha\alpha^\prime}\Gamma^{\alpha\alpha^\prime}
	\right)
\end{align}
where $\Gamma := (d^2-d^{\prime 2})v$ and $\Gamma_{\alpha\alpha^\prime}:=(d^2-d^{\prime 2})v_{\alpha\alpha^\prime}$. Contrary to the analogous formula for the {\bf F} $\to$ {\bf M} reduction, the two potentials do not decouple and the action does not reduce to a form quadratic in field strengths. 

\begin{table}
{\small
\renewcommand{\arraystretch}{1.6} 
\begin{tabular}{|c|c|c|c|c|}
 \hline
{\bf T} &potential(s) &field &constraint & EOM
\cr \hline
	dVM &$A_\alpha = d_\alpha v + d^{\alpha\prime}v_{\alpha \alpha\prime}$
		&$W_\alpha = (d^2+d^{\prime2}) A_\alpha$
		&$d^\alpha W_\alpha + d^{\alpha \prime}W_{\alpha \prime} =0$
		&$d^\alpha W_\alpha - d^{\alpha \prime}W_{\alpha \prime} $
\cr
		&$A_{\alpha\prime} = -d_{\alpha\prime} v - d^\alpha v_{\alpha \alpha\prime}$
		&$W_{\alpha\prime} = (d^2+d^{\prime2}) A_{\alpha\prime}$
		&
		&$d_{\alpha \prime}W_\alpha - d_\alpha  W_{\alpha \prime}$
\cr \hline		
	TM &$v=d^{\alpha\prime} \eta_{\alpha\prime} - d^\alpha \eta_\alpha $
		&$G = (d^2 - d^{\prime 2})v$ 
		&$(d^2 + d^{\prime 2})G = 0$ 
		&$d_\alpha G ~,~ d_{\alpha^\prime} G$
\cr  \hline
	SM &$U$ 
		& $\phi = (d^2 + d^{\prime 2}) U$ 
		&$(d^2 - d^{\prime 2}) \phi=0$ 
		&$(d^2 + d^{\prime 2})\phi $
\cr \hline 
	VM &$\Gamma = (d^2-d^{\prime2})V$
		&$W_\alpha = -d_\alpha \Gamma$
		&$d^\alpha W_\alpha - d^{\alpha \prime}W_{\alpha \prime} =0 $
		&$
			d^\alpha W_\alpha + d^{\alpha^\prime} W_{\alpha^\prime}$
\cr
		&
		&$W_{\alpha\prime} = d_{\alpha\prime} \Gamma$ 
		&$d_{\alpha \prime}W_\alpha - d_\alpha  W_{\alpha \prime} =0 $
		&
\cr \hline
\end{tabular}
} 
\caption{{\bf F} $\to$ {\bf T} reduction.}
\label{T:T}
\end{table}

Once more, we find that the reduction {\bf F} $\to$ {\bf T} produces a table with features not present in the na\"ive table derived from {\bf T}. In this case the mismatch is attributable to the reduction of certain irreducible {\bf F} multiplets into representations with more than one prepotential, as required by T-parity. A related feature is that we do not have an interpretation as a de Rham complex in this case ({\it i.e.} table \ref{T:T} does not have the same de Rham-type structure as tables \ref{T:F}, \ref{T:M}, \ref{T:M2}, and \ref{T:T1}).

\subsubsection{T-dual Supergravity}
The reduction of the supergravity fields is as in reference \cite{Polacek:2014cva}: $H_{\vev{\alpha \beta}} \to H_{\alpha \beta^\prime}, H_3=\tfrac12 H^\gamma{}_\gamma = - \tfrac12 H^{\gamma^\prime}{}_{\gamma^\prime}$ with gauge transformations 
\begin{align}
\delta H_{\alpha \alpha^\prime} = d_{\alpha^\prime} L_\alpha - d_\alpha L_{\alpha^\prime} 
~~~\mathrm{and}~~~
\delta H_3 = \tfrac12( d^{\alpha} L_\alpha - d^{\alpha^\prime} L_{\alpha^\prime} ).
\end{align}
The field equations (\ref{E:SuperEinsteinF}) reduce to 
\begin{align}
G_{\alpha \alpha^\prime} &= 
	{a_0} d^2 d^{\prime 2} H_{\alpha \alpha^\prime} 
	-{a_1} (d^2 +d^{\prime 2})(
		p_\alpha{}^\beta H_{\beta \alpha^\prime} 
		+ p_{\alpha^\prime}{}^{\beta^\prime} H_{\alpha \beta^\prime})
\\ &
	-{a_2} (d^2 -d^{\prime 2})(
		p_\alpha{}^\beta H_{\beta \alpha^\prime} 
		- p_{\alpha^\prime}{}^{\beta^\prime} H_{\alpha \beta^\prime})
	-\tfrac{a_2}2 (d^2-d^{\prime 2}) d_\alpha d_{\alpha^\prime} H^\gamma{}_\gamma
	+2a_3 p_\alpha{}^\beta p_{\alpha^\prime}{}^{\beta^\prime} H_{\beta \beta^\prime}
\cr  
G^\alpha{}_\alpha &= 
		{a_0} d^2 d^{\prime 2} H^\alpha{}_\alpha
	+a_2(d^2-d^{\prime 2}) d_\alpha d_{\alpha^\prime} H^{\alpha \alpha^\prime} 
	+a_3 (p^2  + p^{\prime ^2})H^\alpha{}_\alpha 
\nonumber	
\end{align}
Substituting the values $(a_0 , a_1, a_2, a_3)= (1,-\tfrac12, \tfrac12, \tfrac12)a_0$, we get
\begin{align}
\tfrac1{a_0}G_{\alpha \alpha^\prime} &= 
	d^2 d^{\prime 2} H_{\alpha \alpha^\prime} 
	+ p_{\alpha^\prime}{}^{\beta^\prime} d^2 H_{\alpha \beta^\prime}
	+ p_\alpha{}^\beta d^{\prime 2} H_{\beta \alpha^\prime} 
	-\tfrac14 (d^2-d^{\prime 2}) d_\alpha d_{\alpha^\prime} H^\gamma{}_\gamma
	+ p_\alpha{}^\beta p_{\alpha^\prime}{}^{\beta^\prime} H_{\beta \beta^\prime}
\cr 
\tfrac1{a_0}G^\alpha{}_\alpha &= 
	-\tfrac12 (d^2 - d^{\prime 2})^2 H^\alpha{}_\alpha
	+\tfrac12(d^2-d^{\prime 2}) d_\alpha d_{\alpha^\prime} H^{\alpha \alpha^\prime} .
\end{align}
Equivalently, we can reduce the {\bf F} action (\ref{E:Faction}) to obtain the {\bf T} action
\begin{align}
S_{\bf T} = a_0 \int d^{3}x \int d^2\theta d^2\theta^\prime \, 
	\Big\{
			H^{\alpha \alpha^\prime} d^2d^{\prime2}  H_{\alpha \alpha^\prime} 
			- p_\alpha{}^\beta H^{\alpha \alpha^\prime}  d^{\prime2}  H_{\beta \alpha^\prime} 
			- p_{\alpha^\prime}{}^{\beta^\prime} H^{\alpha \alpha^\prime}  d^2H_{\alpha \beta^\prime} 
\cr 	
		- p_\alpha{}^\beta H^{\alpha \alpha^\prime} p_{\alpha^\prime}{}^{\beta^\prime} H_{\beta \beta^\prime} 
		 +d_\alpha d_{\alpha^\prime} H^{\alpha \alpha^\prime} (d^2 - d^{\prime 2}) H_3
		- 2 H_3 (d^2 - d^{\prime 2})^2 H_3 
	\Big\} .
\nonumber
\end{align}

These results are to be compared to those of reference \cite{Polacek:2014cva}. Unfortunately, this is complicated due to an error in the algebra that qualitatively changes the form of the equations of motion in the current version of that paper.

\subsection{M,T \texorpdfstring{$\to$}{\textrightarrow} S} 

Finally, reduction to {\bf S} can be implemented through the {\bf M} branch or the {\bf T} branch. 
In the {\bf M} $\to$ {\bf S} reduction, this is simply dimensional reduction along the 3-direction.
In the {\bf T} $\to$ {\bf S} branch, the reduction is performed by solving the F-constraint (``sectioning'') which sets $p^\prime = p$.
Either way, the result is a $(2+1)$-dimensional theory with an $N=2$ supersymmetry algebra 
\begin{align} 
\{ d_\alpha , d_\beta \} = 2 p_{\alpha \beta}
,~~~
\{ d_{\alpha} , {d}_{\beta^\prime} \} = 0
,~~~
\{ {d}_{\alpha^\prime} , {d}_{\beta^\prime} \} = 2 p_{\alpha^\prime \beta^\prime} .
\end{align}
More convenient, however, is to go to a complex basis in {\bf T} analogous to that of {\bf M}. Then {\bf M} $\to$ {\bf S} and {\bf T} $\to$ {\bf S} reduction can be treated simultaneously as both give the 3D, $N=2$ algebra
\begin{align}
\{ d_\alpha , d_\beta \} = 0
,~~~
\{ d_{\alpha} , \bar {d}_{\beta} \} = p_{\alpha\beta} 
,~~~
\{ \bar {d}_{\alpha} ,\bar {d}_{\beta} \} = 0,
\end{align}
gotten from {\bf M} by dropping the dot ${}_{\dot \alpha} \to {}_\alpha$ and from {\bf T} by making complex combinations proportional to $d_\alpha \pm i  d_{\alpha^\prime}$. 

Since the reduction {\bf M} $\to$ {\bf S} is trivial, we choose to reduce {\bf T} instead. The combinations $d^2\pm d^{\prime 2}$ reduce to $d^2 + \bar d^2$ and $2i d^\alpha \bar d_\alpha$ respectively. Thus, the scalar multiplet $\phi \to d^2 U + \bar d^2 U$ becomes a 3D, $N=2$ chiral field $\Phi = d^2 U$ with real prepotential in the combination invariant under shifts $\Phi \to \Phi + i c$ by imaginary constant scalars. On the other hand, the tensor multiplet becomes an $N=2$ vector multiplet $G \to 2i d^\alpha \bar d_\alpha v$. 
That this is a vector multiplet, can be seen by reducing the vector multiplet itself: First, we make the complex combinations $W_\alpha = W_\alpha + i W_{\alpha^\prime}$ and its conjugate. Then, $W_\alpha = -d_\alpha \Gamma$ with $\Gamma = 2i  d^\alpha \bar d_\alpha V$ invariant under the gauge transformation $\delta V= \Omega + \bar \Omega$ by chiral scalars $\bar d_\alpha \Omega = 0$. Thus, the fieldstrength of the three-dimensional, $N=2$ vector multiplet simplifies to a real-linear superfield $\Gamma = \bar \Gamma$ and $\bar d^2 \Gamma = 0$. 
The fields of the dual vector multiplet reduce similarly.

\section{Conclusions}

In this paper, we have studied the superspace formulation of toy F-theories ({\bf F}) in the linearized limit. 
Such superspaces reduce to analogues of M-theories ({\bf M}), and type-II theories with manifest T-duality ({\bf T}).
These reduce in turn to a type-II ``string'' theory ({\bf S}) in its usual formulation.
Focusing on the case in which {\bf S} corresponds to a three-dimensional string, we constructed the {\bf F} superspace to which it lifts. Field theories in {\bf F} are {\it a priori} ten-dimensional but are subject to the F-constraint (\ref{E:Fconstraint}) effectively reducing them to four dimensions. The matter and gravitational theories were constructed in the quadratic approximation. All of the matter theories have abelian gauge invariances and severely restricted interactions. The F-constraint follows from the gauge invariance of the gauge invariance in {\bf F}-gravity. 

Reduction {\bf F} $\to$ {\bf M} is achieved by solving the F-constraint. The reduction of the matter and gravitational theories was worked out in detail. Due to the gauge invariance of the {\bf F}-theory matter fields, the reduced matter fields are modified versions of the usual matter multiplets familiar from 4D, $N=1$ superspace. 
In the same vein, {\bf F} gravity reduces to old minimal supergravity but with a real prepotential for the chiral compensator, which representation contains a gauge 3-form. 

By contrast, {\bf F} reduces to {\bf T} by dimensional reduction along a space-like direction. The resulting theory has Spin(2,2) $\cong$ Spin(2,1)$\times$Spin(1,2) symmetry corresponding to the manifest T-duality invariance of the underlying type-II theories. Again, the reduced matter representations are significantly more complicated than what one would na\"ively expect. The supergravity reduction produces a result that is expected to agree with that of reference \cite{Polacek:2014cva} after correction of an error in that paper. 

In light of these results, there are many interesting extensions of this work to pursue. For example, we have not yet attempted to extend the analysis to the case of the D=3 superstring {\it \`a la} \cite{Siegel:1993xq, Siegel:1993th, Siegel:1993bj} nor have we studied the first quantization of these theories.
It is of interest to understand the results of this work in the context of such a quantization, possibly in the presence of membranes.
Alternatively, it should be straightforward to extend this construction to the more interesting dimensions corresponding to D=4 and D=6 strings. 

In a different direction, the extension to the non-linear theory would be an interesting avenue for further investigation into this {\bf F} theory for D=3. The non-linear extension of the supergravity theory is complicated by the lack of chirality in {\bf F}. Even for the matter theories, the non-abelian extension will have to contend with the presence of forms of degrees higher than 1 in the generic multiplet. (Although the 1-form does not have this problem, there we are unable to impose conventional constraints on the non-abelian connection.) A related line of investigation is a proper study of the {\bf F}-theory forms. Throughout this work, we have encountered ``{\bf F}-orms'' that reduce to ordinary $p$-forms upon reduction to {\bf M} and {\bf T} but these are not $p$-forms in the ten-dimensional sense since pairs of SO(2,3) indices are symmetric rather than anti-symmetric under interchange. Understanding how the latter (and their {\bf T}-theory analogues) arise and their relation to the former may shed light on the (super-)geometry of {\bf F}-theory.

\section*{Acknowledgements}
W{\sc dl}3 thanks the Simons Center for Geometry and Physics for hospitality during the {\sc xii} Simons Workshop at which time this project was started. W{\sc dl}3 is partially supported by the U{\sc mcp} Center for String \& Particle Theory.

This work is supported in part by National Science Foundation grants
PHY-1316617, 
PHY-0652983, and PHY-0354401. 

\appendix

\section{\texorpdfstring{\hskip-.4em}{} is for ``Algebra"}
\label{S:A}
\subsection{SL(4; {\bf R})}

The easiest way to derive identities for the {\bf F}-theory spinor algebra is from 4D $©$-matrix algebra:  The steps are:   (1) Derive the algebra for products of $©$-matrices by considering special cases.  Identically valued pairs of indices give Minkowski metrics $ú$, differently valued indices give antisymmerized products of $©$'s.  (The terms are fairly obvious, with signs from anticommuting different $©$'s past each other.)  (2) Replace $©$'s with spinor derivatives $d$, and metrics $ú$ with ``spacetime" derivatives $p=-i»$ (also, 4-vector indices with 4-spinor).  At this point the results are manifestly SL(4) covariant:
\begin{align}
d_Œ d_º =\ & d^2_{Œº} + p_{Œº} \nonumber \\
d_Œ d_º d_© =\ & -·_{Œº©¶}d^{3¶} + (p_{Œº}d_{©} - p_{Œ©}d_º + p_{º©}d_Œ) \nonumber \\
d_Œ d_º d_© d_¶ =\ & -·_{Œº©¶}d^4
	+ (p_{Œº}d^2_{©¶} - p_{Œ©}d^2_{º¶} + p_{Œ¶}d^2_{º©} + p_{º©}d^2_{Œ¶} - p_{º¶}d^2_{Œ©} + p_{©¶}d^2_{Œº}) \nonumber \\
	& + (p_{Œº}p_{©¶}  - p_{Œ©}p_{º¶} + p_{Œ¶}p_{º©}) \nonumber \\
d_Œ d_º d_© d_¶ d_· =\ & -(p_{Œº}·_{©¶·½} - p_{Œ©}·_{º¶·½} + p_{Œ¶}·_{º©·½} - p_{Œ·}·_{º©¶½} + p_{º©}·_{Œ¶·½} \nonumber \\
	& - p_{º¶}·_{Œ©·½} + p_{º·}·_{Œ©¶½} + p_{©¶}·_{Œº·½} - p_{©·}·_{Œº¶½} + p_{¶·}·_{Œº©½}) d^{3½} \nonumber \\
	& + (p_{Œº}p_{©¶} - p_{Œ©}p_{º¶} + p_{Œ¶}p_{º©}) d_· - (p_{Œº}p_{©·} - p_{Œ©}p_{º·} + p_{Œ·}p_{º©}) d_¶ \nonumber \\
	& + (p_{Œº}p_{¶·} - p_{Œ¶}p_{º·} + p_{Œ·}p_{º¶}) d_© - (p_{Œ©}p_{¶·} - p_{Œ¶}p_{©·} + p_{Œ·}p_{©¶}) d_º \nonumber \\
	& + (p_{º©}p_{¶·} - p_{º¶}p_{©·} + p_{º·}p_{©¶}) d_Œ
\end{align}
The conventions are unambiguous when factors are minimized, except for signs implied for $d^3$ and $d^4$ as chosen by index ordering, and the related choice $·_{1234}=·^{1234}=1$.  (The ``$-$" with all $·$'s will be explained below for Sp(4).)  We then have
\begin{align}
Ó d_Œ , d_º Õ = 2p_{Œº}¼,&â[ d_Œ , d_º ] = 2d^2_{Œº} \nonumber \\
Ó d^2_{Œº} , d_© Õ = -2·_{Œº©¶}d^{3¶}¼,&â[ d^2_{Œº} , d_© ] = -2p_{©[Œ}d_{º]} \nonumber \\
Ó d^{3Œ} , d_º Õ = -2p_{º©}Éü·^{©Œ¶·}d^2_{¶·}¼,&â[ d^{3Œ} , d_º ] = 2¶^Œ_º d^4 \nonumber \\
Ó d^4 , d_Œ Õ = 0¼,&â[ d^4 , d_Œ ] = 2p_{Œº}d^{3º} \nonumber \\
Ó d^2_{Œº} , d^2_{©¶} Õ = -2·_{Œº©¶}d^4 - 2p_{©[Œ}p_{º]¶}¼,&â[ d^2_{Œº} , d^2_{©¶} ] = -2p_{[©[Œ}d^2_{º]¶]}
\end{align}
(Multiple (anti)symmetrizations are nested.)

\subsection{Sp(4; {\bf R})}
\label{S:Sp4}
The next step is to separate Sp(4:$\mathbf R$) traces to find the identities relevant to {\bf F}-theory.
Decomposing 

$$ A_{[Œº]} ­ A_{Œº} - A_{ºŒ}¼,âA_{ҌºÔ} ­ A_{[Œº]} - C\hbox{-tr} $$
$$ A_{(ab)} ­ A_{ab} + A_{ba}¼,âA_{ÓabÕ} ­ A_{(ab)} - ú\hbox{-tr} $$

$$ C_{Œº} = - C_{ºŒ} = - C^{Œº} = - (C_{Œº})*¼,âC^{Œ©}C_{º©} = ¶_º^Œ¼;âÆ^Œ ­ C^{Œº}Æ_º¼,âÆ_Œ = Æ^º C_{ºŒ} $$
$$ -·_{Œº©¶} = C_{Œº}C_{©¶} + C_{Œ©}C_{¶º} + C_{Œ¶}C_{º©} $$

$$ p^2 ­ üp^{Œº}p_{Œº}¼,âp^2_{ҌºÔ} ­ üp_{Ҍ}{}^© p_{ºÔ©}¼;âp_Œ{}^© p_{º©} = p^2_{ҌºÔ} - üC_{Œº}p^2 $$
$$ d^2 ­ üd^Œ d_Œ¼,âë_{Œº} ­ üd_{Ҍ}d_{ºÔ}¼;âd_Œ d_º = p_{Œº} + ë_{Œº} - üC_{Œº}d^2 $$

$$ ü·^{Œº©¶}d^2_{©¶} = ë^{Œº} + üC^{Œº}d^2 $$

\begin{align}
\label{dalgebra}
Ó d^2 , d_Œ Õ = 2d^3_Œ¼,&â[ d^2 , d_Œ ] = -2p_Œ{}^º d_º \nonumber \\
Ó ë_{Œº} , d_© Õ = -2C_{©ÒŒ}d^3_{ºÔ}¼,&â[ ë_{Œº} , d_© ] = -2p_{©ÒŒ}d_{ºÔ} \nonumber \\
Ó d^2 , ë_{Œº} Õ = -2p^2_{ҌºÔ}¼,&â[ d^2 , ë_{Œº} ] = 2p_{Ҍ}{}^© ë_{ºÔ©} \nonumber \\
Ó ë_{Œº} , ë_{©¶} Õ = 2C_{ŒÒ©}C_{¶Ôº}d^4 - p_{ҩҌ}p_{ºÔ¶Ô}¼,&â
	[ ë_{Œº} , ë_{©¶} ] = -2p_{ҩҌ}ë_{ºÔ¶Ô} + p_{ҩҌ}C_{ºÔ¶Ô}d^2
\end{align}

\begin{align}
\label{E:d4a}
2d^4 = (d^2)^2 + p^2 = üd^Œ d^2 d_Œ - p^2 \\
\label{E:d4b}
d_{(Œ}d^2 d_{º)} = -2p_{(Œ}{}^© ë_{º)©}¼,âd_{Ҍ}d^2 d_{ºÔ} = 2p^2_{ҌºÔ}\\
\label{E:d5}
d^2 d_Œ d^2 = 2p^2_{ҌºÔ}d^º¼,âüd^Œ(d^2)^2 d_Œ = -p^{2ҌºÔ}ë_{Œº}
\end{align}

\subsection{SL(2; {\bf C})}
\label{S:FtoM}
We now break Sp(4;$\mathbf R$) $\to$ SL(2;$\mathbf C$) = Sp(2;$\mathbf C$) for {\bf M}-theory. 
The Sp(4;$\mathbf R$) index decomposes as $Œ£(Œ,ÀŒ)$, each taking two values.
The spinor derivatives decompose as
$$ d_Œ £ å2(d_Œ , Ðd_ÀŒ)¼,âd^2 £ 2(d^2 + Ðd^2) $$
where the new $d^2 := \tfrac12 d^\alpha d_\alpha$ and $\bar d^2 = (d^2)^\ast$.
The tensors decompose as
$$ C_{Œº} £ \begin{pmatrix}C_{Œº} & 0 \cr 0 & C_{ÀŒÀº} \cr \end{pmatrix}¼,â
	p_{Œº} £ \begin{pmatrix}0 & p_{ŒÀº} \cr p_{ºÀŒ} & 0 \cr\end{pmatrix} $$
A traceless-anti-symmetric tensor $X_{\alpha\beta} = \tfrac12 X_{\langle \alpha \beta\rangle}$ decomposes as
\begin{align}
X_{Œº} £ \begin{pmatrix} -C_{Œº}\f1{å2}X_{-1} & X_{ŒÀº} \cr -X_{ºÀŒ} & C_{ÀŒÀº}\f1{å2}X_{-1} \cr\end{pmatrix}¼,â
	X_{-1} = \f1{å2}X^Œ{}_Œ = -\f1{å2}X^{ÀŒ}{}_{ÀŒ}
\end{align}
This gives 
\begin{align}
ë_{Œº} £
\begin{pmatrix}
	-C_{\alpha \beta} ({d}^2-\bar {d}^2) & [{d}_\alpha , \bar {d}_{˼}] \\
	{[\bar {d}_{ÀŒ},{d}_\beta]} & C_{ÀŒÀº} ({d}^2-\bar {d}^2)
\end{pmatrix}	.
\end{align}

\subsection{Spin(2,2) \texorpdfstring{$\to$}{\textrightarrow} Sp(2; $\mathbf R$)}
A simple split in the range of indices $Œ£Œ,Œ'$, without any new normalization factors, is sufficient to reduce from {\bf F} to {\bf T}-theory, with its Sp(2;$\mathbf R$)$\times$Sp(2;$\mathbf R$) symmetry. Explicitly, 
\begin{align}
X_{\alpha \beta} \to 
	\begin{pmatrix} C_{\alpha \beta} X_{3} & X_{\alpha \beta^\prime} 
		\cr -X_{\alpha \beta^\prime}  & -C_{\alpha^\prime \beta^\prime} X_{3} \cr
		\end{pmatrix}¼,â
	X_{3} = \tfrac 12 X^\alpha{}_\alpha = - \tfrac12 X^{\alpha^\prime}{}_{\alpha^\prime} .
\end{align}
Truncating to one Sp(2;$\mathbf R$) factor gives the reduction to {\bf S}-theory.
The 3D, $N=1$ algebra is given by \cite{Gates:1983nr}
\begin{align}
\{ d_\alpha , d_\beta\} = 2 p_{\alpha \beta} 
\end{align}
with $Œ,º=1,2$ and $p_{\alpha \beta} = p_{\beta \alpha}$. The non-trivial identities can be gotten from those of section \ref{S:Sp4} by setting $d^\prime \to 0$ $\Rightarrow$ $p^\prime ,d^3 , d^4\to 0$:
\begin{align}
d^\alpha d_\beta d_\alpha = 0 
,~~~
d^2 d_\alpha = - d_\alpha d^2 = p_{\alphaº} d^\beta
,~~~
\tfrac12 d^\alpha d^2 d_\alpha = -d^2d^2 = p^2.
\end{align}

\section{\texorpdfstring{\hskip-.4em}{} is for ``Bosonic"}
\label{S:B}
In this appendix, we give a toy version of {\bf F}-theory with no fermions. This is the $N=0$ analogue of the construction in the body of the paper. We begin with bosonic {\bf F}-theory ``$p$-forms''.

\subsection{Forms}
\label{section:F-orms}
In this section, we study the bosonic analogue of the ``$p$-forms'' that arise in {\bf F} theory multiplets. Uncovering the structure of these {\bf F}-orms is not entirely straightforward. Importantly, they are {\em not} de Rham forms in ten dimensions because they arise from the F-constraint
\begin{align}
\partial_{[ab}\partial_{cd]} = 0 
	~\Leftrightarrow~
\partial_{a[b}\partial_{cd]} = 0 ,
\end{align}
rather than the relation $[\partial_{ab}, \partial_{cd}] = 0$ on which the de Rham complex is based.

We start with the following assumptions regarding the potential $A$ of a 0-form: Its gauge transformation is a real constant and its field strength is the gradient $F_{ab} = \partial_{ab} A$ with equation of motion $\partial^{ab}F_{ab}{{}={}} 0$. This field strength solves the Bianchi identity $\partial_{a[b}F_{cd]} = 0$.\footnote{It also solves $\partial_{[ab}F_{c]d} = 0$ but this is equivalent since $\partial_{a[b}F_{cd]} = 0$ $\Rightarrow$ $0=\partial_{[ab}F_{cd]} \propto\partial_{a[b}F_{cd]} - \partial_{[bc}F_{d]a}$. We prefer the form presented in the table as it is more easily compared to the form of the equations of motion of the dual table.} 
With this initial condition, the table of $p$-forms is built by increasing the number of indices on the potential $A$. This gives the following structure:
\begin{align}
\label{E:p}
\begin{array}{clll | l}
~~~p~~~
	&\mathrm{~Bianchi~Id.~} 
	& \mathrm{~Solution~} 
	& \mathrm{~Gauge~inv.~} 
	& \mathrm{~~~EOM~} \cr \hline
0 &\partial_{a[b} F_{cd]} =0& F_{ab} = \partial_{ab}A & \delta A =\mathrm{constant~} & \partial^{ab}F_{ab}{{}={}} 0 \cr
1 &\partial_{a[b} F_{cde]} =0& F_{abc} = \partial_{[ab}A_{c]}  & \delta A_a = \partial_{ba} \lambda^b & \partial^{bc}F_{abc}{{}={}} 0 \cr
2 &\partial_{a[b} F_{cdef]} =0& F_{abcd} = \partial_{[ab}A_{cd]}  & \delta A_{ab} = \partial_{c[a} \lambda^c_{b]} & \partial^{cd}F_{abcd}{{}={}} 0 \cr
3 &\mathrm{~~~~~\,none}& F_{abcde} = \partial_{[ab}A_{cde]}  & \delta A_{abc} = \partial_{d[a} \lambda^d_{bc]} &\partial^{de}F_{abcde}{{}={}} 0 
\end{array}	
\end{align}
at which point the table terminates. 
The gauge transformations are unusual in that their gauge parameters are vector-valued; there is no particular symmetry on indices at different heights. 
Note 
that the field strengths for the gauge $p$-form have an unfamiliar ``degree'' (the number of indices is not $p+1$).\footnote{For example, the gauge 1-form is represented as having a 3-form field strength. We can put it in the slightly more familiar form by writing $F_{abc} = \tfrac12 \epsilon_{abc}{}^{de} F_{de}$, in terms of which $F_{ab}= \tfrac12 \epsilon_{ab}{}^{cde}\partial_{de} A_e$, satisfies the Bianchi identity $\partial^{ab}F_{bc} = 0$ and the equation of motion $\partial_{[ab}F_{cd]} {{}={}} 0$. Note that, although we are using $\epsilon$, $F_{ab}$ is {\em not} the Poincar\'e dual of $F_{abc}$.
} 

Table (\ref{E:p}) is not invariant under ``Poincar\'e'' duality. Instead, mapping $F\to \ast F$ and flipping Bianchi identity and equation of motion gives a second table of dual forms
\begin{align}
\label{E:*p}
\begin{array}{clll | l}
~~~p~~~
	&\mathrm{~Bianchi~Id.~} 
	& \mathrm{~Solution~} 
	& \mathrm{~Gauge~inv.~} 
	& \mathrm{~~~EOM~} \cr \hline
\ast 3 &\partial_{ab} F =0 & F=\mathrm{constant~}& \mathrm{~~~none} & \mathrm{~~~none} \cr
\ast 2 &\partial_{[ab} F_{c]}=0 & F_a = \partial_{ab}A^b & \delta A^c = \partial_{ab} \lambda^{[abc]} & \partial^{ab}F_b{{}={}} 0 \cr
\ast 1&\partial_{[ab} F_{cd]} =0& F_{ab} = \partial_{c[a}A^c_{b]} & \delta A_a^c = \partial_{ab}\lambda^{bc}-\tfrac15\delta^c_a \partial\cdot \lambda & \partial^{ab}F_{bc}{{}={}} 0 \cr
\ast 0 &\partial_{[ab} F_{cde]} =0& F_{abc} = \partial_{d[a}A^d_{bc]}  & \delta A_{ab}^c = \partial_{ab}\lambda^c & \partial^{ab}F_{bcd}{{}={}} 0 
\end{array}	
\end{align}
Again, there is no special symmetry implied on indices unless explicitly indicated. For example, the gauge parameter $\lambda^{ab}$ has both symmetric and anti-symmetric parts. (The subtraction of the divergence can be imposed directly on the parameter. Alternatively, the field strength can be defined as $F_{ab} = \tfrac12 \partial_{[ca}A^c_{b]}$ in which case the parameter is completely unconstrained.)

Comparing the tables, we see that the gauge transformation of a $p$-form is of the same structure as the field strength of a $\ast (3-p)$-form. 
The actions that imply the equations of motion are all of the form $\int F^2$ with the appropriate number of indices on $F$ contracted with the (inverse) Minkowski metric $\eta^{ab}$. 

\subsubsection{Reduction to {\bf M}}

The reduction of a gauge $p$-form gives an {\bf M}-theory $p$-form and nothing else. The part of the field strength $F_{ab\dots} = \partial_{[ab}A_{\dots]}$ with $p+2$ components vanishes identically since $\partial_{ab} \to 0$ in {\bf M}. The ${}_{-1}$ component, on the other hand becomes an exact $(p+1)$-form field strength for a gauge $p$-form. This form satisfies a Maxwell-type equation on-shell, as given in the last column of (\ref{E:p}).

The reduction of a $\ast p$-form gives an {\bf M}-theory $(2-p)$-form and an auxiliary $(p-1)$-form. The $(2-p)$-form is {\bf M}-closed and satisfies a Maxwell type equation ({\it i.e.}~is {\bf M}-co-closed) on-shell. The other component of the $\ast p$-form equation of motion reduces to $\partial_{-1a}F_{-1bc\dots} {{}={}} 0$ which implies that the $(p-1)$-form is constant on-shell.

\subsection{F-gravity}
\label{S:BosonicFgravity}
As a warm-up to the derivation of the linearized supergravity action in {\bf F}-theory, we work out the analogous calculations in a bosonic model containing only a symmetric, traceless rank-2 tensor $h_{ab}$. 
The functional derivative of the quadratic action $S=\int L(h)$ defines the linearized Einstein tensor $G(h) := \delta S/ \delta h$. This is required to be gauge invariant $G(\delta h) = 0$ and we can use this fact to derive $G(h)$.

Up to surface terms, the variation of the action can be written as
\begin{align}
\int \delta L {{}={}} \int \vev{h \delta G},
\end{align}
where, in this appendix only, $\vev{M} = M^a{}_a$ will stand for the (bosonic) trace ({\it e.g.} $\vev{h} = 0$).
The linearized Einstein tensor is
\begin{align}
G = \left\{ a_0 \vev{\partial \partial} h + a_1 \partial \partial h -a_2 \partial h \partial \right\}, 
\end{align}
with $\{\dots\}$ standing for twice the symmetric and traceless part ({\it e.g.} $\{h\} = 2 h$).
Here and henceforth we use matrix notation but all $\partial$s act on the field. In this notation, the gauge transformation is
\begin{align}
\delta h & = \left\{ \partial \lambda\right\}
	= \partial \lambda +\lambda \partial - \tfrac25 \vev{\partial \lambda} 
\end{align}
Then
\begin{align}
2 \delta G &= \{  2a_0 \vev{\partial \partial} \partial \lambda 
	+ a_1 \partial \partial \partial \lambda 
		+ a_1 \partial \partial \lambda \partial
		- \tfrac25 a_1 \partial \partial \vev{\partial \lambda}  \cr
&\hspace{3cm}	-a_2 \partial \partial \lambda \partial 
		-a_2 \partial \lambda \partial \partial
		+\tfrac25 a_2 \partial \partial \vev{\partial \lambda}
	\}\cr
&= 2a_0 \vev{\partial^2} \{ \partial \lambda \}
	+ a_1 \{ \partial^3 \lambda \}
	-a_2 \{ \partial \lambda \partial^2 \} \cr
&\hspace{3cm}	+ (a_1 - a_2) \{ \partial^2 \lambda \partial \}
	- \tfrac25 (a_1-a_2) \{\partial^2 \} \vev{\partial \lambda} 
\end{align}
but $[\partial]= 2\partial$ and $[\lambda]=2 \lambda$ imply that $\partial^2 \lambda \partial = \partial\lambda \partial^2$ so this simplifies to
\begin{align}
2\delta G &=2a_0 \vev{\partial^2} \{ \partial \lambda \}
	+ a_1 \{ \partial^3 \lambda \}
	+ (a_1 -2 a_2) \{ \partial^2 \lambda \partial \}
	- \tfrac25 (a_1-a_2) \{\partial^2 \} \vev{\partial \lambda} .
\end{align}
The F-constraint $\partial_{[ab}\partial_{cd]} = 0$ implies $\partial_{[ab}\partial_{cd]} \partial^{de} = 0$.
Expanding this out,
multiplying by $h$ on the left and $\lambda$ on the right, and contracting all indices in the two inequivalent ways gives the two equations
\begin{align}
\vev{h \vev{\partial^2} \partial \lambda}  = 2 \vev{h \partial^3 \lambda}
~~~\mathrm{and}~~~
\vev{h\partial^2}\vev{\lambda \partial} = 2\vev{h \partial^2\lambda\partial} .
\end{align}
With this, the variation of the action collapses to 
\begin{align}
2\delta L &{{}={}} 2a_0 \vev{ h \vev{\partial^2} \partial \lambda }
	+ a_1 \vev{h \partial^3 \lambda }
	+ (a_1 -2 a_2) \vev{ h \partial^2 \lambda \partial }
	- \tfrac25 (a_1-a_2) \vev{h \partial^2 } \vev{\partial \lambda} \cr
&=(4a_0+  a_1) \vev{h \partial^3 \lambda }
	+ \tfrac15 (a_1 -6a_2) \vev{ h \partial^2 \lambda \partial }.
\end{align}
Therefore, taking $(a_1,a_2) = (- 4, - \frac{2}{3})a_0$, the bosonic {\bf F} action is invariant.

\subsubsection{Reduction to {\bf M}}
Expanding out the {\bf F}-theory graviton and its gauge transformation, we find
\begin{align}
h_{ab} \to 
	\left( 
		\begin{array}{cc}
			h_{-1-1} & h_{-1 b}\\
			h_{-1 a} & h_{ab}
		\end{array}
	\right)
~~~\mathrm{and}~~~
\delta h_{ab} \to 
	\left( 
		\begin{array}{cc}
			-\tfrac65 \partial_c \xi^c & \partial^c \xi_{bc}\\
			\partial^c \xi_{ac} & \partial_{(a} \xi_{b)} -\tfrac45 \eta_{ab} \partial_c \xi^c
		\end{array}
	\right)
\end{align}
where we defined $\xi_a:= -\xi_{-1a}$. Note that $h_{-1-1}= h^c{}_c$. Define the combinations
\begin{align}
g_{ab} := h_{ab} -\tfrac23 \eta_{ab} h^c{}_c
~,~~
C_{abc} := \epsilon_{abc}{}^d h_{-1d}
~~~\Rightarrow~~~
\delta g_{ab} = \partial_{(a}\xi_{b)}
~,~~~
\delta C_{abc} = \partial_{[a}\lambda_{bc]}
\end{align}
where $\lambda_{ab} : = - \tfrac14 \epsilon_{ab}{}^{cd} \xi_{cd}$. The 4-form field strength $F_{abcd}:= \tfrac1{3!} \partial_{[a} C_{bcd]}$ is related to $\partial^ah_{-1a} = - F_{0123}$. 
Then the action becomes 
\begin{align}
S_{\mathbf M} &=  \int d^4 x \Big\{
	2a_0 g^{ab}\Box g_{ab} 
	-a_1\partial_a g^{ab}\partial^c g_{bc}
	-\tfrac25 (2a_1 + 3a_2) g\partial^a\partial^b g_{ab}
\cr &\hspace{4cm}
	+\tfrac1{25}(10a_0 +13a_1 +12 a_2) g\Box g 
	+(a_1+2a_2) (\partial^a h_{-1a})^2
	\Big\} .
\cr
&= 4a_0 \int d^4 x \Big\{
	-\tfrac12 (\partial_c g_{ab})^2
	+(\partial^b g_{ab})^2
	+\tfrac12 (\partial_a g)^2
	+g\partial^a\partial^b g_{ab}
	+\tfrac{1}{18} F_{abcd}F^{abcd}
	\Big\} .
\end{align}
When $a_0=-\tfrac18$, this reproduces the linearized action for Einstein gravity and a decoupled gauge 3-form with 4-form field strength $F= dC$ (which carries no four-dimensional degrees of freedom). We conclude that this quadratic bosonic {\bf F}-theory reproduces the linearized limit of the generalized geometry construction proposed in \cite{Berman:2010is}.


{\small \bibliography{/Users/wdlinch3/Dropbox/Rashoumon/BibTex}} 

\bibliographystyle{unsrt}

\end{document}